\documentclass[twoside]{article}
 
    \input{ArticlePreamble.sty}
    
    \input{Qcircuit}
    
\author{Juan Bermejo-Vega\footnote{juan.bermejovega@mpq.mpg.de}, Maarten Van den Nest\footnote{maarten.vandennest@mpq.mpg.de} \\ \\ {\normalsize Max-Planck-Institut f\"ur Quantenoptik,} \\ {\normalsize Hans-Kopfermann-Stra\ss{e}  1, 85748 Garching, Germany. }}
\title{\vspace{-2cm} Classical simulations of Abelian-group normalizer\\ circuits  with intermediate measurements}

\begin{document}

\maketitle

\begin{abstract}
Quantum normalizer circuits were recently introduced as generalizations of Clifford circuits \cite{VDNest_12_QFTs}: a \textit{normalizer circuit} over a finite Abelian group $G$ is composed of the quantum Fourier transform (QFT) over $G$, together with gates which compute quadratic functions and automorphisms. In \cite{VDNest_12_QFTs} it was shown that every normalizer circuit can be simulated efficiently classically. This result provides a nontrivial example of a family of quantum circuits that \emph{cannot} yield exponential speed-ups in spite of usage of the QFT, the latter being a central quantum algorithmic primitive. Here we extend the aforementioned result in several ways. Most importantly, we show that normalizer circuits supplemented with intermediate measurements can also be simulated efficiently classically, even when the computation proceeds \emph{adaptively}. This yields a generalization of the Gottesman-Knill theorem (valid for $n$-qubit Clifford operations \cite{Gottesman_PhD_Thesis, gottesman_knill}) to quantum circuits described by arbitrary finite Abelian groups. Moreover, our simulations are twofold:  we present efficient classical algorithms to sample the measurement probability distribution of any adaptive-normalizer computation, as well as to compute the amplitudes of the state vector in every step of it. Finally we develop a generalization of the \emph{stabilizer formalism} \cite{Gottesman_PhD_Thesis, gottesman_knill} relative to arbitrary finite Abelian groups: for example we characterize how to update stabilizers under generalized Pauli measurements and provide a normal form of the amplitudes of generalized stabilizer states using quadratic functions and subgroup cosets.
\end{abstract}

\section{Introduction}

 Investigating the power of restricted families of quantum circuits is a fruitful approach to understanding how the power of quantum computers compares to that of classical ones.  A celebrated result in this respect is the Gottesman-Knill theorem, which states that any quantum circuit built out of Clifford gates (Hadamards, CNOTs, $\uppi/2$-phase gates) and Pauli measurements can be efficiently simulated on a classical computer \cite{Gottesman_PhD_Thesis, gottesman_knill, nielsen_chuang}; thus, a quantum computer that works exclusively with these operations cannot achieve \textit{exponential quantum speed-ups}.

The Gottesman-Knill theorem illustrates how subtle the frontier between classical and quantum computational power can be. For example, even though Clifford circuits can be simulated efficiently classically, replacing the $\uppi/2$-phase gates by a $\uppi/4$-phase gate immediately yields a quantum \emph{universal} gate set \cite{Boykin_etal_99_Clifford_pifourth_is_universal_OPEN,Boykin_etal_00_Clifford_pifourth_is_universal_ELSEVIER}. Another interesting feature is that, even though the computing power of Clifford circuits is not stronger than classical computation,  their behavior is genuinely quantum: they can be used, for instance, to prepare highly entangled states (such as cluster states \cite{raussen_briegel_01_Cluster_State, nest06Entanglement_in_Graph_States, raussen_briegel_onewayQC}), or to perform quantum teleportation \cite{gottesman_knill}. Yet, in spite of the high degrees of entanglement that may be involved, the evolution of a physical system under Clifford operations can be tracked efficiently using a Heisenberg picture: the \textit{stabilizer formalism}, a fundamental tool in quantum error correction \cite{Gottesman_PhD_Thesis, gottesman_knill, nielsen_chuang}.

In this work we study the computational power of \textit{normalizer circuits}. These circuits were introduced in \cite{VDNest_12_QFTs} by one of us, as a class of quantum computations  generalizing the Clifford circuits, as well as  standard extensions of the latter to qudits \cite{Gottesman98Fault_Tolerant_QC_HigherDimensions, dehaene_demoor_hostens}. A normalizer circuit over a finite Abelian group $G$ is a quantum circuit comprising unitary gates that implement the quantum Fourier transform (QFT) over $G$, quadratic functions of the group and automorphisms. If $G$ is chosen to be $\mathbb{Z}_2^{n}$ (i.e.\ the group of $n$-bit strings with addition modulo 2), normalizer circuits precisely coincide with the unitary Clifford circuits (i.e.\ those composed of CNOT, $H$ and $\uppi/2$ phase gates). In \cite{VDNest_12_QFTs} it was shown that arbitrary normalizer circuits (acting on computational basis states and followed by computational basis measurements) can be simulated classically efficiently.

An interesting feature of the normalizer circuit formalism is the presence of QFTs over any finite Abelian group. Of particular interest is the group $\mathbb{Z}_{2^n}$, since its corresponding QFT is the ``standard'' quantum Fourier transform \cite{nielsen_chuang}; which lies at the core of several famous quantum algorithms, such as  factoring and computing discrete logarithms \cite{Shor}. More generally, QFTs over Abelian groups are central ingredients of quantum algorithms to find hidden subgroups of Abelian groups \cite{lomont_HSP_review,childs_lecture_8, childs_vandam_10_qu_algorithms_algebraic_problems}.  In contrast with the role of QFTs in quantum speed-ups,  the normalizer circuit formalism provides a nontrivial example of a family of quantum computations that \emph{cannot} yield exponential speed-ups, in spite of usage of the QFT.

Here we further extend the classical simulation results of  \cite{VDNest_12_QFTs}. We do so by considering normalizer circuits where \emph{intermediate measurements} are allowed at arbitrary times in the computation---whereas in \cite{VDNest_12_QFTs} only terminal measurements were considered. More precisely we define \textit{adaptive normalizer circuits} over $G$ to comprise the following three fundamental ingredients:
\begin{itemize}
\item \textbf{Normalizer gates} over $G$, i.e. QFTs, automorphism gates, quadratic phase gates.
\item \textbf{Measurements} of generalized Pauli operators over $G$ at arbitrary times in the computation.
\item \textbf{Adaptiveness:} the choice of normalizer gate at any time may depend (in a polynomial-time computable way) on the outcomes obtained in all previous measurement rounds.
\end{itemize}
If $G$ is chosen to be $\mathbb{Z}_2^{n}$, the corresponding class of adaptive normalizer circuits precisely corresponds to the class of adaptive Clifford circuits allowed in the original Gottesman-Knill theorem.

This paper contains several results, summarized as follows:
\begin{itemize}
\item[I.] \textbf{A Gottesman-Knill theorem for all finite Abelian groups} (Theorem \ref{thm_main}). Given any Abelian group $G$, every poly-size adaptive normalizer circuit over $G$, acting on any standard basis input, can be efficiently simulated by a classical computer. That is, we show that the conditional probability distribution arising at each measurement (given the outcomes of the previous ones) can be sampled in classical polynomial time.

\item[II.]  \textbf{A stabilizer formalism for finite Abelian groups}. Generalizing the well-known stabilizer formalism for qubits, we develop a stabilizer formalism for arbitrary Abelian groups. This framework is a key ingredient to efficiently track the evolution of quantum states under normalizer circuits. In particular, our results are:
    \begin{itemize}
    \item We provide an analytic formula, as well as an efficient algorithm,  to compute the dimension of any stabilizer code over a finite Abelian group (Theorem \ref{thm structure test}).
    \item We provide an analytic formula, as well as an efficient algorithm, to compute the update of any stabilizer group under Pauli measurements over arbitrary finite Abelian groups (Theorem \ref{thm_Measurement_Update_rules}).
    \end{itemize}
\item[III.] \textbf{A normal form for stabilizer states} (Theorem \ref{thm Normal form of an stabilizer state}). We give an analytic formula to characterize the amplitudes of stabilizer states over Abelian groups and show how to compute these amplitudes efficiently. It follows that all stabilizer states over Abelian groups belong to the class of Computationally Tractable (CT) states, introduced in \cite{nest_weak_simulations}. The interest in this property is that all CT states can be simulated classically in various contexts well beyond the setting of the present work---cf. \cite{nest_weak_simulations} for a discussion.
\end{itemize}
In all the results above the term \textit{efficient} is used as synonym of ``in polynomial time in $\log{|G|}$'' (where $|G|$ denotes the cardinality of the group $G$). All algorithms presented show good performance regarding computational errors: the sampling algorithm given in theorem \ref{thm_main} is \textit{exact} (i.e.\ it samples the output probability of the adaptive normalizer circuit \emph{exactly} in polynomial time\footnote{In our model, for simplicity we assume availability of a subroutine which allows to generate, with zero error, a uniformly random integer in the interval $[0, N]$ in polylog$(N)$ time, for any integer $N$. Under this assumption, our classical sampling algorithm for simulating normalizer circuits also has perfect accuracy i.e. no additional errors are introduced.}), whereas the algorithms in theorem \ref{thm Normal form of an stabilizer state} yield \textit{exponentially} accurate estimates of state amplitudes and normalization constants.

An important technical difference (and difficulty) compared to the original Gottesman-Knill theorem is that in the context of arbitrary finite Abelian groups (such as $G=\mathbb{Z}_{2^n}$) arithmetic is generally over large integers. This is in contrast to $\mathbb{Z}_2^n$ where arithmetic is simply over $\mathbb{Z}_2$ i.e.\ modulo 2. The difference is in fact twofold:\\

\noindent
$\bullet$\hspace{1mm} First, $\mathbb{Z}_2$ is a \textbf{field}. As a result, it is possible to describe the ``standard'' stabilizer formalism for qubits with vector space techniques over $\mathbb{Z}_2$. In this context methods like Gaussian elimination have straightforward analogues, which can be exploited in the design of classical algorithms. General Abelian groups are however no longer fields. This complicates both the analytic and algorithmic aspects of our Abelian-group stabilizer formalism due to, for instance, the presence of zero divisors.\\

\noindent $\bullet$\hspace{1mm} Second, in $\mathbb{Z}_2$ arithmetic is with small numbers (namely 0s and 1s), whereas in general finite Abelian groups arithmetic is with \textbf{large integers}. For example, this is the case with $G=\mathbb{Z}_{2^n}$. Of course, one must beware that some problems in number theory are widely believed to be \textit{intractable} for classical computers: consider, for instance, the integer factorization problem or computing discrete logarithms. One of the main challenges in our scenario is to show that the ``integer arithmetic'' used in our classical simulation algorithms can be carried out efficiently. For this purpose, a significant technical portion of our work is dedicated to solving \textit{systems of linear equations modulo a finite Abelian group}, defined as follows:  given a pair of finite Abelian groups  $G_{sol}$ and  $G$ (both of which are given as a direct product of cyclic groups), and a homomorphism $\mathcal{A}$ between them, we look at systems of the form $\mathcal{A}(x)=b$ where $x\in G_{sol}$ and $b\in G$. We present polynomial-time deterministic classical algorithms for counting and finding solutions of these systems. These efficient algorithms lie at the core of our classical simulations of normalizer circuits.\\

Finally, we mention that the stabilizer formalism has been used in a variety of settings (both for qubits and $d$-level systems) beyond the context of the Gottesman-Knill theorem. This includes e.g.\ 	measurement-based quantum computation \cite{raussen_briegel_onewayQC,ZhouZengXuSun03,Schlingemann04ClusterStates}, quantum error-correction  and fault-tolerance \cite{Gottesman98Fault_Tolerant_QC_HigherDimensions, BravyiKitaev05MagicStateDistillation,CampbellAnwarBrowne12MagicStateDistillation_QUTRITS, CampbellAnwarBrowne12MagicStateDistillation_in_all_prime_dimensions}, secret-sharing \cite{Hillery99Quantum_Secret_Sharing, Cleve99Quantum_Secret_Sharing, Gottesman00Quantum_Secret_Sharing}, topological systems \cite{kitaev_anyons, BombinDelgado07HomologicalQEC, BullockBrennen07QUDIT_surface_code,DuclosCianci_Poulin13ToricCode_QUDITS} and other applications. The mathematical tools developed in the present work may therefore also have applications outside the realm of classical simulations of quantum circuits.

\subsection{Relation to previous work}

 In \cite{VDNest_12_QFTs} it was proven that one can sample classically in poly-time the output distribution of any \emph{non-adaptive} normalizer circuit followed by a terminal measurement in the standard basis. Our work extends  this result in various ways, as outlined above in I-II-III. Main differences are the fact that here we consider adaptive normalizer circuits, and two different types of simulations: sampling output distributions and computation of amplitudes.

To our knowledge, ref.\ \cite{VDNest_12_QFTs} and the present work are the first studies to investigate normalizer circuits over arbitrary finite Abelian groups, including those of the form $G=\mathbb{Z}_{d}^{m}$ where $d$ can be an \textit{exponentially} large number, such as $d=2^n$; they are also the first to consider normalizer operations that act on high-dimensional physical systems without a natural tensor product decomposition (such as \ $\C^{p}$ where $p> 2^{n}$ is an exponentially big prime number), or clusters of heterogeneous qudits (e.g. $\C^{a}\times \C^{b}\times\C^{c}$ when $a, b, c$ are different, as opposed to $\C^{d^{\otimes n}}$).

Restricting to groups of the form $G=\mathbb{Z}_{d}^{m}$ where $d$ is \textit{constant}, our work recovers previous results regarding classical simulations of Clifford circuits for \textit{qudits}. We emphasize that in this second scenario $d$ is a \textit{fixed} parameter that does not scale; this is in contrast with the cases studied in \cite{VDNest_12_QFTs} and in the present paper. We briefly summarize prior work on qudits.
 \begin{itemize}
\item Results when $d$ is a \textbf{constant prime} number: if $d = 2$, the ability to \textit{sample} classically efficiently follows from the Gottesman-Knill theorem \cite{Gottesman_PhD_Thesis,gottesman_knill}, whereas the computation of \textit{amplitudes} from \cite{dehaene_demoor_coefficients};  for prime values of $d$ larger than 2, techniques given in \cite{Gottesman98Fault_Tolerant_QC_HigherDimensions} yield efficient sampling simulations also for adaptive Clifford circuits.
\item Results when $d$ is an \textbf{arbitrary constant}: techniques given in  \cite{dehaene_demoor_hostens} can be used to simulate \textit{non}-adaptive Clifford circuits  followed by a terminal standard basis measurement (sampling output distributions and computation of amplitudes); tools developed in  \cite{deBeaudrap12_linearised_stabiliser_formalism} can be used to sample in the adaptive case.
\end{itemize}
Finally, our work also connects to previous studies on the simulability of Abelian quantum Fourier transforms (QFTs), such as \cite{aharonov_AQFT, yoran_short_QFT, brown_QFT}. The relation between normalizer circuits to the quantum circuits considered in those works was discussed in \cite{VDNest_12_QFTs}.

\section{Outline of the Paper}

 This paper is organized as follows.

Section \ref{sect_Summary_of_Concepts} summarizes the key concepts of this work: Pauli and Clifford operators, and normalizer circuits. Sections \ref{sect_prelimin_fab} and \ref{section Pauli and Clifford and Normalizer} contain technical preliminaries. Section \ref{sect_prelimin_fab} gives an introduction to the theory of finite Abelian groups, including a number of efficient classical algorithms to solve algebraic computational problems: the methods presented therein form the basic technical machinery used this work. Section \ref{section Pauli and Clifford and Normalizer} gives a detailed account of the mathematical properties of Pauli, Clifford and (unitary) normalizer operations.

The remaining sections contain the main results of our work. In section \ref{section Abelian group stabilizer formalism}, a theory of Abelian-group stabilizer codes is developed. Section \ref{section  Pauli measurements Implementation} explains how intermediate (generalized) Pauli operator measurements can be implemented, and how they transform Abelian-group stabilizer states. In section \ref{section Gottesman-Knill theorem} we show how to simulate adaptive normalizer circuits classically, discuss the power of these operations for state preparation and give normal forms for stabilizer states.

\section{Summary of Concepts \label{sect_Summary_of_Concepts}}

 In this section we introduce the most prominent quantum-mechanical concepts that feature in this article, namely,  the notions of Pauli/Clifford operators and normalizer circuits over \textit{finite Abelian groups}, as defined in \cite{VDNest_12_QFTs}. Our aim is to illustrate the main ideas behind these group-theoretical quantum circuits by presenting, without details, some key definitions with several examples. The latter will also explain how normalizer circuits generalize the notions and Clifford operators for qubits \cite{nielsen_chuang} and qudits \cite{Knill96non-binaryunitary, Gottesman98Fault_Tolerant_QC_HigherDimensions}. The technical aspects of these quantum operations will be postponed to section \ref{section Pauli and Clifford and Normalizer}.

\subsection{The Hilbert space associated with a group \texorpdfstring{$G$}{G}}

Let $\Z_d=\{ 0, 1,\ldots, d-1\}$ be the additive group of integers modulo $d$. Then
\begin{equation}\label{G}
G=\DProd{d}{m}
\end{equation} denotes a finite Abelian group, whose elements are $m$-tuples of the form $g=(g(1),\allowbreak \ldots,\allowbreak g(m))$ with $g(i)\in\Z_{d_i}$. Addition of two group elements is component-wise modulo $d_i$. Every finite Abelian group can be expressed as a product of the type (\ref{G}) via isomorphism, yet computing this decomposition is regarded as a difficult computational problem\footnote{The problem is at least as hard as factoring integers, since decomposing $G=\Z_N^{\times}$ yields an efficient algorithm to compute the Euler Totient function: the latter can be used to factorize in polynomial time (see e.g.\ \cite{Shoup08_A_Computational_Introducttion_to_Number_Theory_and_Algebra} chapter 10). Efficient \textit{quantum} algorithms to decompose Abelian groups exist, at least for ``reasonably presented'' groups: this was shown in \cite{mosca_phd, cheung_mosca_01_decomp_abelian_groups} for black-box finite Abelian groups with unique encodings.}; throughout this paper,  a product decomposition (\ref{G}) of $G$ is always explicitly given. The order (or cardinality) of $G$ is denoted by $\mathfrak{g}$, and fulfills $\mathfrak{g}=d_1 d_2 \cdots d_m$.

Any group $G$ as in (\ref{G}) is naturally associated to a $\mathfrak{g}$-dimensional Hilbert space  $\C^G=\C^{d_1}\otimes\cdots\otimes \C^{d_m}$ with a basis $\mc{B}$ labeled by group elements
\begin{equation}\label{Computational Basis EQUATION}
\ket{g}=\ket{g(1)}\otimes\cdots\otimes \ket{g(m)} \quad \textnormal{for all} \quad g\in G.
\end{equation}
This basis is henceforth called the \textit{standard basis}.

\subsection{Pauli operators}

 We now introduce the following operators acting on $\C^G$:
\begin{equation}\label{Pauli Operators DEFINITION}
X(g) := \sum_{h\in G} |h+g\rangle\langle h|,\qquad \qquad Z({g}) := \sum_{h\in G} \chi_{  g}({  h}) |{  h}\rangle\langle h|.
\end{equation}
Here  $g\in G$ and $\chi_g$ is a homomorphism\footnote{That is, $\chi_g$ fulfills $\chi_g(h+h')=\chi_g(h)\chi_g(h')$ for every $h$, $h'\in G$.} from $G$ into the multiplicative group of nonzero complex numbers $\C^{*}$, defined as
\begin{equation}\label{Character Functions DEFINITION}
\chi_g(h)=\exp{\left(2\pii \sum_{i=1}^{m} \frac{g(i)h(i)}{d_i}\right)};
\end{equation}
the functions $\chi_g$ are known as the \textit{character functions} of $G$. All operators  $X(g)$, $Z(g)$ are \textit{unitary}. Indeed, $X(g)$ acts as a permutation on the standard basis; second, since $\chi_g(h)$ is a complex phase, the operator $Z(g)$ is unitary as well.

With definitions (\ref{Pauli Operators DEFINITION}--\ref{Character Functions DEFINITION}), a Pauli operator over $G$ (hereafter often simply denoted \emph{Pauli operator}) is any unitary operator of the form
\begin{equation}\label{sigma}
\sigma(a,g,h) := \gamma^a Z(g)X(h),
\end{equation}
where $\gamma:= \euler^{\imun\uppi /\mathfrak{g}}$ is a primitive root of unity, and $a\in\mathbb{Z}_{2\mathfrak{g}}$.  The triple $(a,g,h)$ describing the Pauli operator is called the \textit{label} of $\sigma$. It is important to observe that, although $\sigma$ is a $\mathfrak{g}\times\mathfrak{g}$ matrix, its label $(a,g,h)$ is  an \textit{efficient}  description of itself comprising $O(\log\mathfrak{g})$ bits; from now on, we will specify Pauli operators in terms of their labels, and refer to the latter as the \textit{standard encoding} of these operators.

It was  proved in \cite{VDNest_12_QFTs} that the set $\mathcal{P}^{G}$ of all Pauli operators over $G$ forms a (finite) group, which we call the \textit{Pauli group} (over $G$); this is reviewed in section~\ref{section Pauli and Clifford and Normalizer}.

\subsection{Clifford operators and normalizer circuits}

 A unitary operator $U$ on $\C^G$ is called a \textit{Clifford operator} (over $G$)   if $U$ maps the Pauli group ${\cal P}^G$ onto itself under the conjugation map $\sigma\to U\sigma U^{\dagger}$. It is easy to see that the set of all Clifford operators forms a group, called the Clifford group ${\cal C}^G$. Formally speaking, the Clifford group is the normalizer of the Pauli group in the full unitary group acting on $\C^G$.

Next we define three basic classes of unitary operators on $\C^G$ which are known to belong to the Clifford group \cite{VDNest_12_QFTs}.

\begin{itemize}
\item[{\bf I.}] {\bf Quantum Fourier transforms.}
The quantum Fourier transform (QFT) over $\mathbb{ Z}_{d_i}$ is the following unitary operator on $\C^{d_i}$: \be\label{QFT_cyclic} F_i =\frac{1}{\sqrt{d_i}} \sum \omega_i^{xy} |x\rangle\langle y| ; \quad\quad  \omega_i = \exp\left(\frac{2\pii}{d_i}\right) \ee where the sum is over all $x, y\in \mathbb{ Z}_{d_i}$. The QFT over the entire group $G$ is given by $F = F_1\otimes \cdots \otimes F_m$, which acts on the entire space $\C^G$. Any operator obtained by replacing a subset of operators $F_i$ in this tensor product by identity operators is called a partial QFT.

\item[{\bf II.}]{\bf Automorphism gates.} Given an automorphism $\alpha$ of $G$,  the associated automorphism gate $U_{\alpha}$ maps $|{  g}\rangle\to |\alpha({  g})\rangle$.

\item[{\bf III.}] {\bf Quadratic phase gates.} A function $\xi$ from $G$ to the  nonzero complex numbers is called \textit{quadratic} if there exists a bilinear\footnote{$B$ is \textit{bilinear} if it is a character in both arguments: i.e. $B(g+g',h)=B(g,h)B(g',h)$ and $B(g,h+h')=B(g,h)B(g,h')$. In particular, all characters are bilinear.} function $B$ such that for every ${  g}, h\in G$ it holds that \begin{equation}\label{quadratic_bilinear}\xi({g}+ h) =\xi({g})\xi(h)B(g, h).\end{equation}
Given a quadratic function $\xi$ on $G$, the \textit{quadratic phase gate} $D_{\xi}$ is the diagonal operator mapping $|{  g}\rangle\to \xi({  g})|{  g}\rangle$.  For every quadratic function $\xi$ the complex number $\xi(g)$ fulfills $\xi(g)^{2\mathfrak{g}}=1$ and is therefore a complex phase  \cite{VDNest_12_QFTs};  as a result, every quadratic phase gate is a (diagonal) unitary operator.

The mathematical properties of quadratic functions are reviewed in section \ref{section Quadratic functions}.
\end{itemize}
A unitary operator which is either a (partial) quantum Fourier transform or its inverse, an automorphism gate or a quadratic phase gate is generally  referred to as a \textit{normalizer gate}. A quantum circuit composed entirely of normalizer gates is called a \textit{normalizer circuit over $G$}. The \textit{size} of a normalizer circuit is the number of normalizer gates of which it consists. A full description of every normalizer gate that is part of a normalizer circuit (type, action, number of qubits on which it acts, etc) can be stored  efficiently (with  \polylog{\mathfrak{g}} bits of memory) in a computer, using---what we call---the \emph{standard encodings} of these gates (to be properly defined in section \ref{section Standard encodings of normalizer circuits}); it follows that every \polylog{\mathfrak{g}} normalizer circuit can be described efficiently as a list of normalizer gates.

\subsection{The relationship between normalizer and Clifford operations}

 It was proven in \cite{VDNest_12_QFTs} that every normalizer gate (and circuit) is a Clifford operator, but it is not known whether all possible Clifford operators can be implemented via normalizer gates. Such a question is of considerable relevance, since the finding of a non-normalizer Clifford operation could lead to a new quantum gate. However, the authors believe that such an operation does not exist; we  \textit{conjecture} that any Clifford operator can be implemented as a poly-size normalizer circuit (conjecture \ref{thm_conjecture}). We refer to section \ref{section conjecture} for a discussion and some supporting evidence.

\subsection{Examples \label{sect_examples}}

 Here we give examples of Pauli and normalizer operations for several groups. We illustrate in particular how the definitions of the preceding section generalize existing notions of Pauli and Clifford operators for qubits and qudits.

\subsubsection{\texorpdfstring{$G=\Z_{2}^{m}$}{G = (Z2) to the m}}

In this case the standard definition of $m$-qubit Pauli operators is recovered. To see this, first note that we have $\C^G = \mathbb{C}^2\otimes  \cdots \otimes\mathbb{C}^2$ i.e.\ the Hilbert space is a system of $m$ qubits. Let $\sigma_x$ and $\sigma_z$ denote the standard Pauli matrices and let $g\in \mathbb{Z}_2^m$ be an $m$-bit string. Then, applying definition (\ref{Pauli Operators DEFINITION}) one finds that
\be X(g) = \sigma_x^{g(1)}\otimes \cdots \otimes \sigma_x^{g(m)},\qquad \qquad  Z(g) = \sigma_z^{g(1)}\otimes \cdots \otimes \sigma_z^{g(m)}.\ee
 Here $g\in \mathbb{Z}_2^m$ is an $m$-bit string: i.e.\  $g(i)\in\{0, 1\}$. In short, $X(g)$ is a tensor product of $\sigma_x$-matrices and identities, and $Z(g)$ is a tensor product of $\sigma_z$-matrices and identities. Therefore, every Pauli operator (\ref{Pauli Operators DEFINITION}) has the form $\sigma \propto U_1\otimes \cdots  \otimes U_m$ where each $U_i$ is a single-qubit operator of the form $\sigma_x^u \sigma_z^v$ for some $u, v\in \mathbb{Z}_2$. This recovers the usual notion of a Pauli operator on $m$ qubits \cite{nielsen_chuang}.

It was shown in \cite{VDNest_12_QFTs} that the CNOT gate and the CZ gate (acting between any two qubits), the Hadamard gate and the Phase gate $S=$ diag$(1, i)$ (acting on any single qubit) are examples of normalizer gates for $G=\Z_{2}^{m}$. Note that these gates are the standard building blocks of the well known class of Clifford circuits for qubits. In fact, the \textit{entire} Clifford group for qubits is generated by this elementary gate set \cite{nielsen_chuang}.

\subsubsection{\texorpdfstring{$G=\Z_{d}^{m}$}{G = (Zd) to the m}}\label{Examples Paulis/Normalizer for qudits}

 In this case the Hilbert space $\C^G = \mathbb{C}^d\otimes \cdots \otimes\mathbb{C}^d$ is a system of $m$ $d$-level systems (qudits) and Pauli operators have the form $\sigma \propto U_1\otimes \cdots  \otimes U_m$, where each $U_i$ is a single-qudit operator of the form $X_d^u Z_d^v$ for some $u, v\in \mathbb{Z}_d$. Here  $X_d$ and $Z_d$ are the usual generalizations of $\sigma_x$ and $\sigma_z$, respectively \cite{Knill96non-binaryunitary, Gottesman98Fault_Tolerant_QC_HigherDimensions}. These operators act on a single $d$-level system as follows: \be\label{X_Z_qudits} X_d = \sum |x+1\rangle\langle x| \quad \mbox{ and }\quad Z_d = \sum \euler^{2\pii x/d}|x\rangle\langle x| \ee where the sums run over all $x\in \mathbb{Z}_d$. Examples of normalizer gates over $\mathbb{  Z}_d^m$  are generalizations of the CNOT, CZ,  Hadamard and Phase gates to qudits, as follows:
\begin{eqnarray}\label{gates_qudits}
\mbox{SUM}_d &=& \sum |x, x+ y\rangle\langle x, y|; \\ \mbox{CZ}_d &=& \sum \omega_d^{xy} |x, y\rangle\langle x, y|; \quad \omega_d:= \euler^{2\pii /d} \\ F_d &=&\frac{1}{\sqrt{d}} \sum \euler^{2\pii xy/d} |x\rangle\langle y|;\\  S_d &=& \sum  \xi_d^{x(x + d)}|x\rangle\langle x|; \quad \xi_d:= \euler^{\pii/d}.\end{eqnarray}
Here $x$ and $y$ rum over all elements in $\mathbb{Z}_d$. To show that SUM$_d$ is a normalizer gate, note that $(x, y)\to(x, x+y)$ is indeed an automorphism of $\mathbb{Z}_d\times \mathbb{Z}_d$. The gates $\mbox{CZ}_d$ and $S_d$ are quadratic phase gates; see Ref.\@\cite{VDNest_12_QFTs}. In addition, the ``multiplication gate'' $M_{d, a} = \sum |ax\rangle \langle x|$ is also a normalizer gate, for every $a\in \mathbb{Z}_d$ which is coprime to $d$. Indeed, for such $a$ the map $x\to ax$ is known to be an automorphism of $\mathbb{Z}_d$. It is known that the \textit{entire} Clifford group for qubits (for arbitrary $d$) is generated by the gates SUM$_d$, $F_d$, $S_d$ and $M_a$ \cite{dehaene_demoor_hostens}.

\subsubsection{\texorpdfstring{$G = \mathbb{Z}_{2^m}$}{G = Z mod (2 to the m)}}

 One can also consider $G$ to be a single cyclic group, such as $G= \mathbb{  Z}_{2^m}$. In this case, $\C^G$ is a $2^m$-dimensional Hilbert space with standard basis $\{|0\rangle,\ldots, |2^m-1\rangle\}$. Comparing with the previous examples, the important difference with e.g.\ $\mathbb{Z}_2^m$ is that the structure of $\mathbb{Z}_{2^m}$ does not naturally induce a factorization of the Hilbert space into $m$ single-qubit systems. As a consequence, normalizer gates over $\mathbb{Z}_{2^m}$ act globally on ${\cal H}$, in contrast with the previous examples.

Examples  of normalizer gates are now given by $F_{2^m}$, $S_{2^m}$ and $M_{2^m,\, a}$, following the definitions of the previous example with $d=2^m$. Note that $F_{2^m}$ is the ``standard'' QFT used in e.g Shor's algorithm and the phase estimation quantum algorithm.

\section{Preliminaries on Finite Abelian Groups \label{sect_prelimin_fab}}

 This section is reserved for the main group-theoretical notions used in this work, and their computational aspects. The concepts and algorithms introduced in this section will be essential to construct a theory of stabilizer codes for physical systems of the form $\C^{G}$ (cf. eq. (\ref{Computational Basis EQUATION})), where $G$ can be any finite Abelian group (\ref{G}).

\subsection{Conventions}

 Throughout this section we fix the canonical decomposition $G$ to be $G=\DProd{d}{m}$, as in (\ref{G}); this choice is completely arbitrary. In computational complexity theory a (classical or quantum) algorithm is said to be \textit{efficient} if it solves a given computational problem of input-size $n$ in (classical or quantum) \poly(n) time: when  one looks at problems related to finite Abelian groups, this will be synonym of ``in \ppolylog{|G_1|,|G_2|,\ldots,|G_n|} time'', being $G_1,\ldots,G_n$ the groups involved in a problem of interest (we look at concrete problems in section \ref{sect_CompComp_FiniteAbelianGroups}).

Last, we introduce a set of \textit{canonical generators} $e_i$ of $G$. For every $i$ ranging from 1 to $m$, ${  e}_i=(0,\ldots,1_i,\ldots,0)\in G$ denotes the group element which has $1\in\mathbb{Z}_{d_i}$ in its $i$-th component and zeroes elsewhere, where $0$ in slot $k$ represents the neutral element in $\mathbb{Z}_{d_k}$. The $m$ elements ${  e}_i$ generate $G$,  for any element ${ g}\in G$ can be naturally written as ${  g} = \sum g(i) {  e}_i $, and play a similar role as the canonical basis vectors of vector spaces like $\R^{m}$ or $\C^{m}$ (though $G$ is \textit{not} a vector space).

\subsection{Character and quadratic functions\label{section Quadratic functions}}

 The character functions $\chi_g$ of $G$ (eq.\ \ref{Character Functions DEFINITION}),  form a finite Abelian group with the multiplication $\chi_g\chi_h=\chi_{g+h} $, called the \textit{character group} or dual group $\widehat{G}$; the latter is isomorphic to $G$ via $g\leftrightarrow\chi_g$. Moreover, for every $g, h, h'\in G$ the following equalities hold:
\begin{equation}\label{eq Properties of Character functions}
\text{Linearity:\quad} \chi_g(h+h') = \chi_g(h)\chi(h'); \qquad \text{Label symmetry:\quad} \chi_g(h) = \chi_h(g).
\end{equation}
Relationships (\ref{eq Properties of Character functions}) are useful to manipulate characters symbolically.

All character functions are quadratic functions: indeed, it is trivial to check that function $B(g,h)\equiv 1$  fulfills
\begin{equation}
\label{bilinear} B(g+h, x) = B(g, x)B(h, x)\quad \mbox{and}\quad B(x, g+h) = B(x, g)B(x, h)
\end{equation}
for every $g$, $h$, $x\in G $, so it is bilinear; $B$ can then be used to write every character in the form (\ref{quadratic_bilinear}).
Furthermore, the set of all quadratic functions of $G$  forms a finite Abelian group with the multiplication $\xi \xi'$, which contains the character group $\widehat{G}$ as a subgroup \cite{VDNest_12_QFTs}: finiteness follows from $\xi(g)^{2\mathfrak{g}}=1$; eqs.\ (\ref{quadratic_bilinear},\ref{bilinear}) can be used to check closure, associativity, commutativity and that the inverse of any quadratic function $\xi$ (its complex conjugate $\overline{\xi}$) is again quadratic.

We finish this section with some examples of quadratic functions over finite Abelian groups (more examples can be found in  ref.\@ \cite{VDNest_12_QFTs}).\\

\noindent \textbf{Example 1:} We consider $G=\mathbb{Z}_2^m$. Letting $A$ be an $m\times m$ matrix with entries in $\mathbb{Z}_2$ and letting $a\in \mathbb{Z}_2^m$, the following functions are quadratic:
\be \xi_A: x\to (-1)^{x^TAx}\quad \mbox{ and }\quad \xi_a: x\to \imun^{a^Tx}\ee
where $x\in\mathbb{Z}_2^m$ and $a^Tx$ is computed over $\mathbb{Z}_2$ (i.e.\ modulo 2). Note that the exponent in $\xi_A$ is polynomial of degree 2 in $x$, whereas the exponent in $\xi_a$ has degree 1. To prove that these functions are quadratic in the sense used in this work, we note that
\begin{eqnarray} \xi_A(x+y) &=& \xi_A(x)\xi_A(y) (-1)^{x^T(A+A^T)y} \\\label{i_quadratic} \xi_a(x+y)&=& \xi(x) \xi(y) (-1)^{q(x, y)} \quad \mbox{ with } q(x, y) =  (a^Tx)(a^Ty)\end{eqnarray}
Identity (\ref{i_quadratic}) can be proved by distinguishing between the 4 cases $a^Tx, a^Ty\in\{0, 1\}$. The above identities can be used to show that $\xi_A$ and $\xi_a$ are quadratic.\\

\noindent \textbf{Example 2:} Considering  a single cyclic group $G=\mathbb{Z}_d$, examples of quadratic functions are \be z\to \omega^{{  b}{  z}^2 + cz} \quad \mbox{ and } z\to \gamma^{  b  z(z + d)}; \quad \omega := \euler^{2\uppi \imun/d}, \ \gamma := \omega^{1/2}.\ee We refer to  \cite{VDNest_12_QFTs} for a proof.

\subsection{Orthogonal subgroups}

 The character functions $\chi_g$ give rise to a set-theoretical duality, sometimes called orthogonality of Abelian subgroups (although it differs from the usual orthogonality of vector spaces). Given a subgroup $H$ of the finite Abelian group $G$, its orthogonal subgroup $H^\perp$ is defined as
\begin{equation}\label{orthogonal group EQUATION}
H^\perp = \left\lbrace g\in G \: \colon \: \chi_h(g)=1, \; \textnormal{for all} \; h\in H \right\rbrace.
\end{equation}
Note that $H^{\perp}$ is indeed a subgroup of $G$. The main properties of $H^{\perp}$ are summarized below.
\begin{lemma}[\textbf{Orthogonal subgroup}]\label{properties of orthogonal subgroups}
Consider a finite Abelian group $G$ as in (\ref{G}) and let $H$ and $K$ be two arbitrary subgroups. Then the following statements hold:
\begin{itemize*}
\item[(a)] $(H^{\perp})^{\perp} = H$
\item[(b)] $|H^{\perp}| = |G/H| = \mathfrak{g}/|H|$
\item[(c)] $H\subseteq K$ if and only if $K^{\perp}\subseteq H^{\perp}$
\item[(d)] $(H\cap K)^{\perp} = \langle H^{\perp},\,  K^{\perp}\rangle$
\end{itemize*}
\end{lemma}
\begin{proof}
(a-b) are well-known, see e.g.\ \cite{lomont_HSP_review}. (c) is proved straightforwardly by applying definitions. We prove (d). Since $H\cap K$ is contained in both $H$ and $K$ if follows that $H^{\perp}\subseteq (H\cap K)^{\perp}$ and $K^{\perp}\subseteq (H\cap K)^{\perp}$. Therefore $\langle H^{\perp},\, K^{\perp}\rangle \subseteq (H\cap K)^{\perp}$. We  show $\langle H^{\perp},\, K^{\perp}\rangle^{\perp} \subseteq H\cap K$, which implies the reversed inclusion. This comes from the fact that $g\perp\langle H^{\perp},\, K^{\perp}\rangle$ implies $g\perp H^{\perp}$ \textit{and} $g\perp K^{\perp}$, or equivalently, $g\in H$ \textit{and} $g\in K$.
\end{proof}

\subsection{Homo-, iso- and auto- morphisms \label{section homo iso auto morphisms}}

 Given two finite Abelian groups $H$ and $G$, a group homomorphism from $H$ to $G$ is a map $\mathcal{A}: H \rightarrow G$ that fulfills $\mathcal{A}(g+h) = \mathcal{A}(g) + \mathcal{A}(h)$ for every $g, h\in H$. (In other words, $\mathcal{A}$ is \textit{linear}.) An isomorphism from $H$ to $G$ is an invertible group homomorphism. An automorphism of $G$ is an isomorphism of the form $\alpha:G\to G$, i.e. from a group onto itself. The set of all automorphisms of $G$ forms a group, called the automorphism group.

Throughout this work, group homomorphisms between Abelian groups are to be described in terms of \emph{matrix representations}, which are defined as follows:
\begin{definition}[\textbf{Matrix representation}]
Given a homomorphism $\mathcal{A}:H\rightarrow G$ between groups $H=\DProd{c}{n}$ and $G=\DProd{d}{m}$, an $m\times n$ integer matrix $A$ is said to be a matrix representation of $\mc{A}$ if its columns $a_i$ are elements of $G$, and if it holds that
\begin{equation}\label{Matrix Representation DEFINITIO-EQUATION}
\mc{A}(h)=Ah\pmod{G}, \quad \textnormal{for every  $h\in H$.}
\end{equation}
Conversely, an $m\times n$ integer matrix $A$ is said to define a group homomorphism from $H$ to $G$ if its columns $a_i$ are elements of $G$  and the operation $Ah\pmod{G}$ is a group homomorphism.
\end{definition}
In (\ref{Matrix Representation DEFINITIO-EQUATION}) we introduced some conventions, that we will use: first, the element $h$ is seen as a \textit{column} of integer numbers onto which $A$ acts via matrix multiplication; second, (mod $G$) indicates that multiplications and sums involved in the calculation of $Ah$ are performed within $G$, taking remainders when necessary.

The main properties of matrix representations are now summarized.
\begin{lemma}\label{Matrix Representation PROPOSITION} Every group homomorphism  $\mathcal{A}: H\rightarrow G$ has a matrix representation. Moreover, an $m\times n$ matrix $A$ with columns $a_i\in G$ defines a homomorphism iff its columns fulfill the equations
\begin{equation}\label{Homomorphism condition}
c_i a_i = 0\pmod{G}, \quad\textnormal{for every } i.
\end{equation}
\end{lemma}
\begin{proof}
First, given $\mathcal{A}$, we  show that $A:= [ \mc{A} (e_1),\allowbreak \ldots, \mc{A} (e_n)]$,  where $e_i:=(0,\ldots,1_i,\ldots, 0)$, is a matrix representation. Since every $h\in H$ decomposes as $h=\sum h(i)e_i$,  it follows, using linearity of $\mathcal{A}$, that
\begin{equation}
\mathcal{A}(h)=\mathcal{A}\left(\sum h(i)e_i\right)=\sum h(i)\mathcal{A}(e_i)=Ah\pmod{G}.
\end{equation}
The right implication of the iff comes readily from
\begin{equation}
c_i a_i=c_i A e_i = A (c_ie_i) = A (0) = 0\pmod{G}
\end{equation} For the converse, we let $A$ act on   $g$, $h$ and $g+h$ without taking remainders; we obtain
\begin{equation}\label{xxx proof matrix representation - homomorphism part}
Ag + Ah = \sum [g(i)+h(i)]a_i ,  \qquad\qquad  A(g+h)=\sum (g+h)(i) a_i.
\end{equation}
Recalling associativity of $H$ and $G$, the latter expression shows that $Ah$ defines a function from $H$ to $\Z^{m}$, and, thus, $Ah \pmod{G}$ is a function from $H$ to $G$. Last, it holds for every $i$ that $g(i)+h(i)=q_i c_i+(g+h)(i)$ for some integers  $q_i$ , since (by definition of the group $H$)  $(g+h)(i)$ is the remainder obtained when $g(i)+h(i)$ is divided by $c_i$ ($q_i$ is  the quotient). It follows, subtracting modularly, that $A(g)+A(h)-A({g+h})=\sum q_i c_i a_i = 0 \pmod{G}$ for every $g$, $h$; and, due to,  $A$ defines a linear map.
\end{proof}

\subsection{Computational group theory\label{sect_CompComp_FiniteAbelianGroups}}

 Computational aspects of finite Abelian groups are now discussed; our discourse focuses on a selected catalog of  computational problems relevant to this work and efficient classical algorithms to solve them. Since this section concerns only classical computational complexity, we will tend to omit the epithet \textit{classical} all the way throughout it.

To start with, we introduce some basic notions of computer arithmetic. From now on,  the  \emph{size} of an integer is the number of bits in its binary expansion. Observe that every group (\ref{G}) satisfies the inequalities $2^{m}\leq \Inputsize$ and $d_i \leq \Inputsize$, for every $i$. It follows readily that $m$ is $O(\polylog{\mathfrak{g}})$, and that one needs at most a $\ppolylog{\mathfrak{g}}$  amount of memory  to store an element $g=(g(1),\ldots,g(m))\in G$ (in terms of bits).

We discuss now how to perform some basic operations efficiently within any finite Abelian group (\ref{G}). First, given two integers $a$ and $b$ of size at most $l$, common arithmetic operations can be computed in \ppoly{l} time with elementary algorithms: such as their sum, product, the quotient of $a$ divided by $b$, and the remainder $a\bmod{b}$ \cite{brent_zimmerman10CompArithmetic}. Therefore, given $g, h\in G$, the sum $g+h$  can be obtained in \ppolylog{\Inputsize} time  by computing the $m$ remainders $g(i) + h(i) \bmod{d_i}$. Similarly, given an integer $n$, the element $ng$  can be obtained in  \ppolylog{\Inputsize, n} time  by computing the remainders $ng(i)$ mod $d_i$.

In connection with section \ref{section homo iso auto morphisms}, it follows from the properties just introduced that matrix representations  can be stored using only a polynomial amount of memory, and, moreover, that given the matrix representation $A$ we can efficiently compute $Ah\pmod{G}$. Specifically,  given a matrix representation $A$ of the homomorphism $\mathcal{A}: H \rightarrow G$, we need $\ppolylog{|H|,|G|}$ space to store its columns $a_i$ as tuples of integers, and $\ppolylog{|H|,|G|}$ time to compute the function $\mathcal{A}(h)$.

Periodically, and at crucial stages of this investigation, some advanced algebraic computational problems are bound to arise. The following lemma compiles a list of group theoretical problems that will be relevant to us and can be solved efficiently by classical computers.
\begin{lemma}[\textbf{Algorithms for finite Abelian groups}]\label{lemma Algorithms for finite Abelian groups}
Given $H$, $K$, two subgroups of $G$, and $\{h_i\}$, $\{k_j\}$, polynomial-size generating-sets of them, there exist efficient classical algorithms to solve the following problems deterministically.
\begin{itemize*}
\item[(a)] Decide whether  $b\in G$ belongs to $H$;  if so, find integers $w_i$ such that $b=\sum w_ih_i$.
\item[(b)] Count the number of elements of $H$.
\item[(c)] Find a generating-set of the intersection $H\cap K$.
\item[(d)] Find a generating-set of $H^{\perp}$.
\item[(e)] Given the system of equations $\chi_{h_i}(g)=\gamma^{a_i}$, find elements $(g_0, g_1, \ldots, g_s)$ such that all solutions can be written as linear combinations of the form $g_0+\sum v_i g_i$.
\end{itemize*}
\end{lemma} 
The  proof of the lemma is divided in two parts which are fully detailed  in appendices \ref{Appendix: Algorithms for Finite Abelian Groups}  and \ref{Appendix: System of linear congruences modulo Abelian group}. The rest of this section  describes the high-level structure of the proof.

In short, appendix \ref{Appendix: Algorithms for Finite Abelian Groups} contains a proof of the following statement.
\begin{lemma}\label{LEMMA REDUCTION TO SYSTEMS OF LINEAR EQUATIONS}
Problems (a-e) in lemma \ref{lemma Algorithms for finite Abelian groups} are polynomial-time reducible to either counting or finding solutions of systems of equations of the form $\mathcal{A}(x)=b$; where $\mc{A}$ is a group homomorphism between two (canonically-decomposed) finite Abelian groups, $\mbf{G}_{{sol}}$ and $\mbf{G}$, to which $x$, $b$ respectively belong; given that a matrix representation of $\mathcal{A}$ is provided.
\end{lemma}
Mind that, since $\mathcal{A}$ is a linear map, the set $X_{sol}$ of solutions of such a system is either empty or a coset with the structure
\begin{equation}\label{Structure of the solutions of A(x)=b EQUATION}
X_{sol}=x_0+\ker{\mathcal{A}}.
\end{equation}
By \textit{finding solutions} we refer to the action of finding one particular solution $x_0$ of the system and a (polynomially-large) generating set of $\ker{\mc{A}}$.

To provide an example, we prove now lemma \ref{LEMMA REDUCTION TO SYSTEMS OF LINEAR EQUATIONS} for the problems (d-e)  in lemma \ref{lemma Algorithms for finite Abelian groups}; for the rest of cases, we refer to the appendices.\\

\textbf{Example: [proof of the (d,e)th cases of lemma \ref{lemma Algorithms for finite Abelian groups}]} First of all, notice that problem (d) reduces to (e) by setting all $a_i$ to be $0$---this yields the system (\ref{orthogonal group EQUATION}), whose solutions are the elements of the orthogonal subgroup. Therefore, it will be enough to prove the (e)th case. Moreover, since the equations $\chi_{h_i}(g)=\gamma^{a_i}$ can be fulfilled for some $g\in G$ only if all $\gamma^{a_i}$ are $\Inputsize$th-roots of the unit, this systems can only have solutions if all  $a_i$ are even numbers. As we can determine it efficiently whether these numbers are even, we assume from now on that it is the case.

Now define a tuple of integers $b$ coefficient-wise as $b(i):=a_i/2$; use the later to  re-write $\gamma^{a_i}=\exp{(2\uppi\textnormal{i}\, b(i)/\Inputsize)}$. Also, denote by $H$ the group generated by the elements $h_i$. By letting $\mathfrak{g}$ multiply numerators and denominators of all fractions in (\ref{Character Functions DEFINITION}), the system of complex exponentials $\chi_{h_i}(g)=\gamma^{a_i}$ can be turned into an equivalent  system of  congruences $\sum_j (\mathfrak{g}/d_j)\, h_i(j) g(j)=b(i) \mod \mathfrak{g}$. Finally, by defining a matrix $\Omega$ with coefficients $\Omega(i,j):=({\mathfrak{g}}/{d_j})\, h_i(j)$ the system can be written as
\begin{equation}
\Omega g = b \pmod{\mbf{G}},
\end{equation}
where  $b$ belongs to $\mbf{G}=\Z_\mathfrak{g}^{r}$, being $r$ the number of generators $h_i$, and we look for solutions inside $\mbf{G}_{sol}=G$. Moreover, the coefficients of $\Omega$ fulfill $d_j \Omega(i,j)=0\bmod{\mathfrak{g}}$; hence, condition (\ref{Homomorphism condition}) is met and $\Omega$ defines a homomorphism.

Finally, $\Omega$ can be computed in $O(\polylog{\Inputsize})$ using aforementioned algorithms to multiply and divide integers. It is now routine to check, using the concepts developed thus far, that both $\log{|\mbf{G}_{sol}|}$ and $\log{|\mbf{G}|}$ are $O(\polylog{\Inputsize})$;  as a result,   the input-size of the new problem, as well as the memory needed to store $\Omega$ and $b$, are all $O(\polylog{\Inputsize})$. \qed\\

\textbf{Remark:} Note that the group homomorphism $\omega(g):=\Omega(g)\pmod{\Z_\mathfrak{g}^{r}}$ fulfills $\ker{\omega}=H^{\perp}$. Therefore, if we substitute $H$ with $H':=H^{\perp}$ in the procedure above, given   $s=\ppolylog{\mathfrak{g}}$ generators of $H'$, we would obtain an $s\times m$ integer matrix $\Omega'$ that defines a second group homomorphism $\varpi:G\rightarrow \Z_\mathfrak{g}^{s}$ such that
\begin{equation}\label{Oracle function definition}
\varpi(g)=\Omega'(g)\pmod{G}\qquad\textnormal{and}\qquad\ker\varpi=H'^{\perp}=H.
\end{equation}
As a result, our algorithm to compute generators of $H^{\perp}$ leads also to an efficient method to construct, given any subgroup $H$ of $G$, a homomorphism $\varpi$ whose kernel is $H$. This fact will be used later in the text, in section \ref{section Gottesman-Knill theorem}. \qed\\

\noindent The final ingredient to complete the proof of lemma \ref{lemma Algorithms for finite Abelian groups} is the following theorem.
\begin{theorem}[\textbf{Systems of linear equations over finite Abelian groups}]\label{THM SYSTEM OF LINEAR CONGRUENCES MODULO AN ABELIAN GROUP} Given any element $b$ of the group $G=\DProd{d}{m}$ and any $m\times n$  matrix $A$ which defines a group homomorphism from  $H=\DProd{c}{n}$  to $G$, consider the system of equations $A x = b \pmod{G}$. Then, there exist classical algorithms to solve the following list of problems in \ppolylog{{|H|}, {|G|}} time.
\begin{enumerate*}
\item Decide whether the system admits a solution.
\item Count the number of different solutions of the system.
\item Find $x_0, x_1, \ldots, x_r\in H$ such that all solutions of the system are linear combinations of the form $x_0+\sum k_i x_i$.
\end{enumerate*}
\end{theorem}
The main ideas underlying the proof of theorem \ref{THM SYSTEM OF LINEAR CONGRUENCES MODULO AN ABELIAN GROUP} are as follows: first, we show how to reduce $Ah=b\pmod{G}$ in polynomial-time to an equivalent system of linear congruences modulo $d$, where $d$ is suitably upper-bounded; second, we apply fast algorithms to compute Smith normal forms to tackle the latter problem \cite{storjohann10_phd_thesis}. For details, we refer to appendix \ref{Appendix: System of linear congruences modulo Abelian group}.

\section{Pauli Operators and Normalizer Circuits over Abelian Groups\label{section Pauli and Clifford and Normalizer}}

\subsection{Manipulation of Pauli operators}

 First, note that every Pauli operator factorizes as a tensor product relative to the tensor decomposition of $\C^G$ i.e.\ $\sigma$ can be written as $\sigma = U_1\otimes\cdots\otimes U_m$ where $U_i$ acts on $\mathbb{C}^{d_i}$. This property simplifies several proofs; it can be verified straightforwardly by applying (\ref{Pauli Operators DEFINITION}) and the definition (\ref{Character Functions DEFINITION}) of the characters of $G$.

Basic manipulations of Pauli operators can be carried out transparently by translating them into transformations of their labels: we review now some of these rules. First, the Pauli matrices (\ref{Pauli Operators DEFINITION}) obey the following commutation rules:
\begin{align}
X({  g})X({  h}) &= X({  g}+{  h}) = X({  h})X({  g}) \nonumber\\  \label{commutation relations Pauli operators} Z({  g})Z({  h})&=Z({  g}+{  h})=Z({  h})Z({  g}) \\
Z({  g}) X({  h}) &= \chi_{  g}({  h})X({  h})Z({  g}).\nonumber\end{align}
Combinations of these rules straightforwardly lead to the next two lemmas.
\begin{lemma}[\textbf{Products and powers of Pauli operators \cite{VDNest_12_QFTs}}]\label{pauli_products_powers PROPERTY}
Consider Pauli operators $\sigma$ and $\tau$ and a positive integer $n$. Then  $\sigma\tau$, $\sigma^n$ and $\sigma^{\dagger}$ are also Pauli operators, the labels of which can be computed in \ppolylog{\mathfrak{g}, n} time on input of $n$ and the labels of $\sigma$ and $\tau$. Moreover,  $\sigma^{\dagger}= \sigma^{2\mathfrak{g}-1}$.
\end{lemma}
\begin{lemma}[\textbf{Commutativity}]\label{Commutativity of Pauli Stabilizer Groups PROPOSITION}
Consider two Pauli operators $\sigma(a_1,g_1,h_1)=\sigma_1$ and $\sigma(a_2,g_2,h_2)=\sigma_2$. Then the following statements are equivalent:
\begin{itemize*}
\item[(i)] $\sigma_1$ and $\sigma_2$ commute;
\item[(ii)] $\chi_{g_1}(h_2)=\chi_{g_2}(h_1)$;
\item[(iii)] $x:=(g_1,h_1)$ and $y:=(h_2,-g_2)$ are orthogonal elements of $G\times G$ i.e.\ $\chi_x(y)=1$.
\end{itemize*}
\end{lemma}
Lemma  \ref{pauli_products_powers PROPERTY} implies that the set of all Pauli operators ${\cal P}^G$ over $G$ forms a (finite) group, called the Pauli group (over $G$).

\subsection{Normalizer quantum circuits\label{section Standard encodings of normalizer circuits}}

 Hitherto we have not considered technical aspects of normalizer circuits, such as how to describe normalizer circuits efficiently, or how to compute their action on Pauli operators; we address these questions in this section.

\subsubsection{Describing normalizer operations}

 In this paper we will be interested in classical simulations of \textit{normalizer circuits}. To make meaningful statements about classical simulations one must first specify which \textit{classical descriptions} of normalizer circuits are considered to be available. In the case of Pauli operators over $G$ (which are particular examples of normalizer operations \cite{VDNest_12_QFTs}), we saw that it is possible to describe them using few (\polylog{\mathfrak{g}}) memory resources, by choosing their labels $(a,g,h)$ as standard encodings; this property holds for \textit{all} normalizer gates and---hence---circuits \cite{VDNest_12_QFTs}: all of them admit efficient classical descriptions. This is discussed next.

\begin{itemize}
\item First, a partial quantum Fourier transform  is described by the set of systems $\mathbb{C}_{d_i}$ on which it acts nontrivially.
\item Second, an automorphism gate is described by the \textit{matrix representation} of the associated automorphism (cf.\ section \ref{section homo iso auto morphisms}).
\item Third, let $\xi$ be an arbitrary quadratic function. Since $\xi(g)^{2\mathfrak{g}}=1$,  there exists $n(g)\in\mathbb{Z}_{2\mathfrak{g}}$ such that $\xi(g) = \euler^{\pii n(g)/\mathfrak{g}}$ for every $g\in G$. It was shown in \cite{VDNest_12_QFTs} that the $O(m^2)$ integers $n(e_i)$ and $n(e_i+e_j)$ comprise an efficient description of $\xi$ and, thus, of the associated quadratic phase gate.
\end{itemize}
Henceforth we will assume that all normalizer gates are specified in terms of the descriptions given above, which will be called their \emph{standard encodings}. The standard encoding of each type of gate comprises polylog$(\mathfrak{g})$ bits. The standard encoding of a normalizer circuits is the sequence of classical descriptions of its gates.

\subsubsection{Normalizer equals Clifford?\label{section conjecture}}

The following theorem states that every normalizer gate belongs to the Clifford group, and the action of any normalizer gate on a Pauli operator via conjugation can be described efficiently classically.
\begin{theorem}[{\bf Normalizer gates are Clifford \cite{VDNest_12_QFTs}}]\label{thm_G_circuit_fundamental}
Every normalizer gate is a Clifford operator. Furthermore let $U$ be a normalizer gate specified in terms of its standard classical encoding as above, and let $\sigma$ be a Pauli operator specified in terms of its label; then the label of $U \sigma U^{\dagger}$ can be computed in \ppolylog{\mathfrak{g}} time.
\end{theorem}
It is unknown whether the entire Clifford group can be generated (up to global phase factors) by normalizer gates in full generality. However, it was proven in  \cite{dehaene_demoor_hostens} (see also the examples in section \ref{sect_examples}) that this is indeed the case for groups of the form $G=\mathbb{Z}_{d}^m$ (i.e.\ $m$ qudit systems); more strongly, every Clifford group element (over $\mathbb{Z}_{d}^m$) can be written as a product of at most polylog$(\mathfrak{g})$ such operators. We \textit{conjecture} that this feature holds true for Clifford operators over \textit{arbitrary} finite Abelian groups.
\begin{conjecture}\label{thm_conjecture} Let $G$ be an arbitrary (canonically decomposed) finite Abelian group. Then, up to a global phase, every Clifford operator over $G$ can be written as a product of \ppolylog{\Inputsize} normalizer gates.
\end{conjecture}
Finally, in the following lemma we provide some partial support for this conjecture. We show that  both  automorphism gates and quadratic phase gates have a distinguished role within the Clifford group, characterized  as follows:
\begin{lemma}
Up to a global phase, every Clifford operator which acts on the standard basis as a permutation has the form $X(g)U_{\alpha}$ for some $g\in G$ and some automorphism gate $U_{\alpha}$. Every diagonal Clifford operator is, up to a global phase, a quadratic phase gate.
\end{lemma}
\begin{proof}
The first statement was proved in \cite{VDNest_12_QFTs}. We prove the second statement. Let $D = \sum \xi(g) |g\rangle\langle g|$ be a diagonal unitary operator (so that $|\xi(g)|=1$ for all $g\in G$) in the Clifford group. Without loss of generality we may set $\xi(0)=1$, which can always be ensured by choosing a suitable (irrelevant) overall phase. Then for every $h\in G$, $D$ sends $X(h)$ to a Pauli operator under conjugation. This implies that there exists a complex phase $\gamma(h)$ and group elements  $f_1(h), f_2(h)\in G$ such that \be\label{condition_diag_clifford} D X(h) D^{\dagger} = \gamma(h) X(f_1(h))Z(f_2(h)). \ee Since $D$ is diagonal, it is easy to verify that we must have $f_1(h)=h$ for every $h\in G$. Now consider an arbitrary $g\in G$. Then
\begin{eqnarray}
D X(h) D^{\dagger}|g\rangle &=& \overline\xi(g) \xi(g+h) |g+h\rangle;\label{quadratic_1}\\
\gamma(h) X(h)Z(f_2(h))|g\rangle &=&  \gamma(h) \chi_g(f_2(h)) |g+h\rangle \label{quadratic_2}.
\end{eqnarray}
Condition (\ref{condition_diag_clifford}) implies that (\ref{quadratic_1}) is identical to (\ref{quadratic_2}) for every $g, h\in G$. Choosing $g=0$ and using that $\xi(0)=1$ and $\chi_0(x)=1$ for every $x\in G$ it follows that $\gamma(h)=\xi(h)$. We thus find that \be\xi(g+h) = \xi(g)\xi(h) \chi_g(f_2(h)). \label{condition_diag_clifford2}\ee The function  $B(g, h):=\xi(g+h)\overline \xi(g)\overline\xi(h)$ is manifestly linear in $g$, since $B(g, h) = \chi_g(f_2(h))$. Furthermore by definition $B$ is symmetric in $g$ and $h$. Thus $B$ is also linear in $h$.
\end{proof}

\section{An Abelian-Group Stabilizer Formalism\label{section Abelian group stabilizer formalism}}

 In this section we develop further the stabilizer formalism for finite Abelian groups as started in \cite{VDNest_12_QFTs}. We provide new analytic and algorithmic tools to describe them and analyze their properties. Throughout this section we consider an arbitrary Abelian  group of the form $G=\DProd{d}{m}$.

\subsection{Stabilizer states and codes}

 Let ${\cal S}$ be a subgroup of the Pauli group ${\cal P}^G$. Then ${\cal S}$ is said to be a stabilizer group (over $G$) if there exists a non-zero vector $|\psi\rangle\in \C^G$ which is invariant under all elements in ${\cal S}$ i.e.\ $\sigma|\psi\rangle = |\psi\rangle$ for every $\sigma\in{\cal S}$. The linear subspace ${\cal V}:= \{|\psi\rangle : \sigma|\psi\rangle = |\psi\rangle \mbox{ for all } \sigma\in {\cal S}\}$ is called the stabilizer code associated with ${\cal S}$. If ${\cal V}$ is one-dimensional, its unique element (up to a multiplicative constant) is called the stabilizer state associated with ${\cal S}$.  In this work we will mainly be interested in stabilizer states. Occasionally, however, it will be useful to consider the general setting of stabilizer codes (cf. e.g.\  theorem \ref{thm structure test}).

Note that every stabilizer group ${\cal S}$ is Abelian. To see this, consider a state $|\psi\rangle\neq 0$ which is invariant under the action of all elements in ${\cal S}$ and consider two arbitrary $\sigma, \tau\in {\cal S}$. Then (\ref{commutation relations Pauli operators}) implies that there exists a complex phase $\alpha$ such that $\sigma\tau = \alpha \tau\sigma$.  It follows that $|\psi\rangle = \sigma\tau |\psi\rangle = \alpha\tau\sigma|\psi\rangle = \alpha|\psi\rangle$, where we have used that $\sigma|\psi\rangle= |\psi\rangle = \tau|\psi\rangle$. We thus find that $|\psi\rangle = \alpha|\psi\rangle$ so that $\alpha = 1$  (i.e.\ $\sigma $ and $\tau$ commute).

Conversely, not every Abelian subgroup of the Pauli group is a stabilizer group. A simple counterexample is the group $\{I, -I\}$ where $I$ is the identity operator acting on $\C^G$.

The support of a stabilizer code ${\cal V}$ is the set of all $g\in G$ for which $|g\rangle$ has a  nonzero overlap with ${\cal V}$ i.e.\ there exists $|\psi\rangle\in {\cal V}$ such that $\langle g|\psi\rangle\neq 0$. The support of a stabilizer state $|\phi\rangle$ is simply the set of all $g\in G$ for which $\langle g|\phi\rangle\neq 0$.

\subsection{Label groups\label{sect_label_groups}}

 Let $\mc{S}$ be a stabilizer group over $G$. The diagonal subgroup $\mc{D}$ is the subgroup of $\mc{S}$ formed by its diagonal operators i.e.\ it  consists of all operators in ${\cal S}$ of the form $\gamma^a Z(g)$.
Second, we introduce two subgroups $\mathbb{H}$ and  $\mathbb{D}$ of $G$ called the label groups of $\mc{S}$:
\begin{align}
\label{Label group H}\mathbb{H} &= \{ h\in G \:\colon\: \textnormal{there exists $\gamma^aZ(g)X(h)\in\mc{S}$} \},\\
\label{Label group D}\mathbb{D} &= \{ g\in G \:\colon\: \textnormal{there exists $\gamma^aZ(g)\in\mc{D}$} \},
\end{align}
Using  (\ref{commutation relations Pauli operators}) it is straightforward to verify that $\mathbb{D}$ is indeed a subgroup of $G$. To prove that $\mathbb{H}$ is a subgroup as well, one argues as follows. Let $\sigma$ be a Pauli operator with label $(a, {  g}, {  h})$. We call $g$ the ``$Z$-component'' and $h$ the ``$X$-component'' of $\sigma$. Denote the $X$-component formally by $\varphi(\sigma):= {  h}$.  Then $\mathbb{H}$ is the image of ${\cal S}$ under the map $\varphi$.  The commutation relations (\ref{commutation relations Pauli operators})   yield \be\label{X_component} \varphi(\sigma\tau)= \varphi(\sigma) + \varphi(\tau)\quad\mbox{ for all } \sigma, \tau \in {\cal S}.\ee
This implies that $\varphi$ is a homomorphism from ${\cal S}$ to $G$. It follows that $\mathbb{H}$ is a subgroup of $G$.
\begin{lemma}[\textbf{Label groups}]\label{label groups of an stabilizer DEFINITION}
Let $\mc{S}$ be a stabilizer group and assume that the labels of $k=$ \ppolylog{\mathfrak{g}} generators of $\mc{S}$ are given as an input. Then the label groups of ${\cal S}$ have the following properties:
\begin{enumerate*}
\item[(i)] $\mathbb{H} \subseteq \mathbb{D}^{\perp}$;
\item[(ii)] Generating sets of $\,\mathbb{H}$, $\mathbb{D}$ can be efficiently computed classically;
\item[(iii)] The labels of a generating set of $\mc{D}$  can be efficiently computed classically.
\end{enumerate*}
\end{lemma}
\begin{proof}
Property (i) is a straightforward consequence of the commutation relations given in lemma  \ref{Commutativity of Pauli Stabilizer Groups PROPOSITION} and the definition of orthogonal subgroup (\ref{orthogonal group EQUATION}).
To show property (ii), recall that the map $\varphi$ defined above is a homomorphism from ${\cal S}$ to $G$ with $\mathbb{H}= $ Im$(\varphi)$. Suppose that ${\cal S}$ is generated by $\{\sigma_1, \ldots, \sigma_k\}$.  Then $\mathbb{H}$ is generated by $\{\varphi(\sigma_1), \ldots, \sigma(\sigma_k)\}$: this yields an efficient method to compute generators of $\mathbb{H}$. To prove the second statement of (ii) as well as (iii) requires more work. The argument is a direct generalization of the proof of lemma 9 in \cite{VDNest_12_QFTs} and the reader is referred to this work.
\end{proof}

\subsection{Certificates}

 The main purpose of this section is to provide a criterion to verify when a stabilizer group gives rise to a one-dimensional stabilizer code i.e.\ a stabilizer state. This is accomplished in corollary \ref{corollary_Uniqueness_Test}. To arrive at this statement we first analyze how the the dimension of a general stabilizer code is related the structure of its stabilizer group.
\begin{theorem}[\textbf{Structure Test}]\label{thm structure test}
Let ${\cal S}$ be a stabilizer group with stabilizer code $\mc{V}$. Then there exists $g_0\in G$ such that
\begin{equation}
(i) \textnormal{ supp}(\mc{V})= g_0 + \mathbb{D}^{\perp}, \qquad\qquad (ii) \textnormal{ dim}(\mc{V})  = \frac{|\mathbb{D}^{\perp}|}{|\mathbb{H}|},
\end{equation}
where $\mathbb{H}$, $\mathbb{D}$ are the label subgroups of ${\cal S}$. Furthermore, there exist efficient classical algorithms to compute a representative $g_0$ of the support, a generating set of $\mathbb{D}^{\perp}$ and the dimension $\textnormal{dim}(\mc{V})$.
\end{theorem}
Before proving theorem \ref{thm structure test}, we note that combining property (ii) together with lemma \ref{label groups of an stabilizer DEFINITION}(i) immediately yield:
\begin{corollary}[\textbf{Uniqueness Test}]\label{corollary_Uniqueness_Test}
Let ${\cal S}$ be a stabilizer group with stabilizer code ${\cal V}$. Then  $\mc{V}$ is one-dimensional if and only if  $\mathbb{H}$ and $\mathbb{D}$ are dual orthogonal subgroups: $\mathbb{H}=\mathbb{D}^{\perp}$.
\end{corollary}
Theorem \ref{thm structure test}(ii) also leads to an alternative formula for the dimension of a stabilizer code:
\begin{corollary}\label{corollary dimension of stabilizer code}
The dimension of ${\cal V}$ equals $\mathfrak{g}/|{\cal S}|$.
\end{corollary}
The result in corollary \ref{corollary dimension of stabilizer code} is well known for stabilizer codes over qubits \cite{Gottesman_PhD_Thesis, nielsen_chuang} (i.e.\ where $G = \mathbb{Z}_2^m$ so that $\Inputsize=2^m$) and qudits (where $G=\mathbb{Z}_d^m$) \cite{Gottesman_PhD_Thesis,gheorghiu11Qudit_Stabilisers}.
\begin{proof}{[of corollary \ref{corollary dimension of stabilizer code}]}
Consider the  map $\varphi: {\cal S}\to G$, defined in section \ref{sect_label_groups}, which is a group homomorphism with image $\mathbb{H}$. Furthermore the kernel of $\varphi$ is precisely the diagonal subgroup ${\cal D}$ of $G$. Since $|\mbox{Im } \varphi| = |{\cal S}|/|\mbox{ker }\varphi|$ it follows that $|\mathbb{H}| = |{\cal S}|/|{\cal D}|$. Finally we claim that ${\cal D}$ and $\mathbb{D}$ are isomorphic groups so that $|{\cal D}| = |\mathbb{D}|$. To prove this,  consider the map $\delta: {\cal D}\to \mathbb{D}$ that sends $\sigma = \gamma^a Z(g)$ to $\delta(\sigma)= g$. Using (\ref{commutation relations Pauli operators}) it follows that this map is a homomorphism; furthermore, it is a surjective one by definition of $\mathbb{D}$, and thus $\textnormal{im}\delta =\mathbb{D}$. The kernel of $\delta$ is the set of all $\sigma\in {\cal S}$ having the form $\sigma = \gamma^a I$. But the only operator in ${\cal S} $ proportional to the identity is the identity itself, since otherwise ${\cal S}$ cannot have a common $+1$ eigenstate. This shows that the kernel of $\delta$ is trivial, so that ${\cal D}$ and $\mathbb{D}$ are isomorphic, as claimed. The resulting identity $|\mathbb{H}| = |{\cal S}|/|\mathbb{D}|$ together with $|\mathbb{D}^{\perp}| = |G|/|\mathbb{D}|$ (recall lemma \ref{properties of orthogonal subgroups}) and theorem \ref{thm structure test}(ii) proves the result.
\end{proof}
We now prove theorem \ref{thm structure test} using  techniques developed in \cite{nest_MMS} where the properties of so-called M-spaces were studied. We briefly recall basic concepts and results.

A unitary operator acting on $\C^G$ is said to be \emph{monomial} if it can be written as a product $U = DP$ where $D$ is diagonal and $P$ is a permutation matrix. A subspace ${\cal M}$ of $\C^G$  is called an \emph{M-space} if there exists a group of monomial unitary operators ${\cal G}$ such that $|\varphi\rangle\in {\cal M}$ iff $U|\varphi\rangle = |\varphi\rangle$ for every $U\in {
\cal G}$. The group ${\cal G}$ is called a stabilizer group of ${\cal M}$. If ${\cal M}$ is one-dimensional, its unique (up to a multiplicative factor) element $|\psi\rangle$ is called an M-state.  The support of ${\cal M}$ is defined analogously to the support of a stabilizer code i.e.\ it is the set of all $g\in G$ such that $|g\rangle$ has a nontrivial overlap with ${\cal M}$. With this terminology, every stabilizer code is an instance of an M-space and every stabilizer state is an M-state. To see this, note that every Pauli operator $\sigma(a, g, h)$ is a monomial unitary operator. Indeed, $\sigma$ can be written as a product $\sigma = DP$ where $D = \gamma^aZ(g)$ is diagonal and $P = X(h)$ is a permutation matrix.

We introduce some further terminology. Let ${\cal G}$ be an arbitrary monomial stabilizer group. For every $g\in G$, let $\mc{G}_{g}$ be the subset of $\mc{G}$ consisting of all $U\in \mc{G}$ satisfying $U|g\rangle\propto |g\rangle$ i.e.\ $U$ acts trivially on $g$, up to an overall phase. This subset is easily seen to be a subgroup of $\mc{G}$. Also, we define the orbit $\mc{O}_g$ of $g$ as:
\begin{equation}\label{orbit DEFINITION} \mc{O}_g = \{h: \exists U\in\mc{G}\mbox{ s.t. } U|g\rangle\propto|h\rangle\}
\end{equation}
In the following  result the support of any M-space is characterized in terms of the orbits ${\cal O}_g$ and the subgroups ${\cal G}_g$.
\begin{theorem}[\textbf{Support of M-space \cite{nest_MMS}}]\label{thm_support_M_space}
Consider an M-space ${\cal M}$ with monomial stabilizer group $\mc{G}$. Then the following statements hold:
\begin{itemize*}
\item[(i)] There exist orbits $\mc{O}_{g_1},\ldots,\mc{O}_{g_\mbf{d}}$ such that $\mbf{d}= \dim({\cal M})$ and \be \textnormal{supp}({\cal M})= {\cal O}_{g_1}\cup\cdots\cup {\cal O}_{g_\mbf{d}}.\ee
\item[(ii)] Consider $g\in G$ and an arbitrary set of generators $\{V_1, \ldots,V_r\}$ of $\mc{G}_g$. Then  $g\in$ supp(${\cal M}$) if and only if $V_{ i}|g\rangle = |g\rangle$ for every $i$.
\end{itemize*}
\end{theorem}
Using this result, we can now prove theorem \ref{thm structure test}.
\begin{proof}\textbf{[of theorem \ref{thm structure test}]}
We apply theorem \ref{thm_support_M_space} to the Pauli stabilizer group $\mc{S}$. In this case, the group $\mc{S}_g$ and the orbit $\mc{O}_g$ fulfill
\begin{equation}\label{Action Orbits&Stabilizer for Pauli Stabilizer Groups}
\mc{O}_g=g+\mathbb{H}, \qquad\qquad \qquad \mc{S}_g = \mc{D}.
\end{equation}
To demonstrate the first identity in  (\ref{Action Orbits&Stabilizer for Pauli Stabilizer Groups}), we use  (\ref{Pauli Operators DEFINITION}) which implies $\sigma(a,x,y)\ket{g}\propto\ket{g+y}$ for every $\sigma(a,x,y)\in\mc{S}$. To show the second identity, first note that $D|g\rangle\propto|g\rangle$ for every diagonal operator $D\in {\cal D}$, showing that ${\cal D}\subseteq {\cal S}_g$. Conversely, if $\sigma\in{\cal S}_g$  has label $(a, x, y)$ then $\sigma |g\rangle\propto |g+y\rangle$. Since $\sigma\in {\cal S}_g$ the state $|g\rangle$ is an eigenvector of $\sigma$; this can only be true if $y=0$, showing that $\sigma\in {\cal D}$.

Using lemma \ref{label groups of an stabilizer DEFINITION}, we can efficiently compute  the labels of a generating set $\{\sigma_1,\ldots,\sigma_r\}$  of $\mc{S}_g=\mc{D}$, where $\sigma_i = \gamma^{a_i} Z(g_i)$ for some $a_i\in\mathbb{Z}_{2\mathfrak{g}}$ and $g_i\in G$. Owing to theorem \ref{thm_support_M_space}(ii), any $g\in G$  belongs to the support of ${\cal V}$ if and only if $\sigma_i\ket{g} = \ket{g}$ for every $i=1,\ldots, r$. Equivalently, $g$ satisfies
\begin{equation}\label{Support of a Stabilizer Code Characteristic EQUATIONS}
\gamma^{a_i}\chi_{g_i}(g)=1 \quad \textnormal{for all} \; i =1,\ldots, r.
\end{equation}
This system of equations is of the type considered in section \ref{sect_CompComp_FiniteAbelianGroups} (cf. lemma \ref{LEMMA REDUCTION TO SYSTEMS OF LINEAR EQUATIONS} and the example after it)  and can, therefore, be transformed into an equivalent linear system over groups: $g\in \textnormal{supp}(\mc{V})\Leftrightarrow \allowbreak\Omega g=b\pmod{\Z_\mathfrak{g}^{r}}$. The elements $g_i$ generate the label group $\mathbb{D}$, and thus the homomorphism $\omega$ defined by $\Omega$ satisfies $\ker{\omega}=\mathbb{D}^{\perp}$. Since the system is linear, its solutions form a coset of the form $\textnormal{supp}(\mc{V})=g_0+\ker{\omega}=g_0+\mathbb{D}^{\perp}$, for some particular solution $g_0$. This shows statement (i).

Further, we combine (i) with theorem \ref{thm_support_M_space}(i) to get a short proof of (ii): the equation
\begin{equation*}
\textnormal{supp}(\mc{V})= \mc{O}_{g_1}\cup\cdots\cup \mc{O}_{g_{\mbf{d}}}=\left(g_1+\mathbb{H}\right)\cup\cdots\cup (g_{\mbf{d}}+\mathbb{H})= g_0+\mathbb{D}^{\perp}
\end{equation*}
implies, computing the cardinalities of the sets involved, that $\textbf{d}|\mathbb{H}|=\dim{\mc{V}}|\mathbb{H}|=|\mathbb{D}^{\perp}|$.

Finally, the ability to compute $g_0$ and to find generators of $\mathbb{D}^{\perp}$ efficiently classically follows applying theorem \ref{THM SYSTEM OF LINEAR CONGRUENCES MODULO AN ABELIAN GROUP} to a linear system  described by a $r\times m$ matrix  $\Omega$ that defines a  homomorphism from $G$ to $\Z_\mathfrak{g}^{r}$, with  $r\in O(\polylog{\Inputsize})$. Furthermore, we can compute $\dim \mc{V}$ directly using formula (ii) together with lemma \ref{label groups of an stabilizer DEFINITION} and the algorithms of lemma \ref{lemma Algorithms for finite Abelian groups}.
\end{proof}

\section{Pauli Measurements in the Stabilizer Formalism}

\subsection{Definition}

 Associated with every Pauli operator $\sigma$ 
(\ref{sigma}) we will consider a quantum measurement in the eigenbasis of $\sigma$. Consider the spectral decomposition $\sigma = \sum \lambda P_{\lambda}$ where $\lambda$ are the distinct eigenvalues of $\sigma$ and $P_{\lambda}$ is the projector on the eigenspace associated with eigenvalue $\lambda$.
Given a state $|\psi\rangle\in \C^G$, the measurement associated with $\sigma$ is now defined as follows: the possible outcomes of the measurement are labeled by the eigenvalues $\{{\lambda}\}$ where each $\lambda$ occurs with probability $\|P_{\lambda}|\psi\rangle\|^2$; furthermore, if the outcome $\lambda$ occurs, the state after the measurement  equals to $P_{\lambda}|\psi\rangle$ up to normalization.

Consider a group $G$ of the form (1), with associated physical system $\C^{G}=\C^{d_1}\otimes\cdots\otimes\C^{d_m}$. We remark that a measurement of the $i$-th system $\mathbb{C}^{d_i}$ in the standard basis $\{|0\rangle, \ldots, |d_i-1\rangle\}$ can be realized as a measurement of a suitable Pauli operator, for every $i$ ranging from 1 to $m$. To keep notation simple, we demonstrate this statement for the special case $G=\Z_{d}^{m}$, yet the argument generalizes straightforwardly to arbitrary $G$. Denote by ${e}_i\in G$ the group element which has $1\in\mathbb{Z}_{d}$ in its $i$-th component and zeroes elsewhere. Then definition (\ref{Pauli Operators DEFINITION}) implies that the Pauli operator $Z(e_i)$ acts as $Z_{d}$ on the $i$-th qudit and as the identity elsewhere, where $Z_d$ was defined in (\ref{X_Z_qudits}). Note that $Z_d$ has $d$ distinct eigenvalues, each having a rank-one eigenprojector $|x\rangle\langle x|$ with $x\in\mathbb{Z}_d$. It follows straightforwardly that measurement of $Z(e_i)$ corresponds to measurement of the $i$-th qudit in the standard basis.

\subsection{Implementation\label{section  Pauli measurements Implementation}}

 It is easily verified that every Pauli operator $\sigma$ can be realized as a polynomial size (unitary) quantum circuit\cite{VDNest_12_QFTs}. Therefore, measurement of $\sigma$ can be implemented efficiently on a quantum computer using standard phase estimation methods \cite{nielsen_chuang}.  Here we provide an alternate method. In particular we show that every Pauli measurement can be implemented using \emph{only} normalizer circuits and measurements in the standard basis. This property will be a useful ingredient in our proof of theorem \ref{thm_main}.
\begin{lemma}[\textbf{\cite{Nielsen02UniversalSimulations}}] \label{lem_Nielsen02UniversalSimulations} For any dimension $d$ and for integers $j$ and $k$ such that $j,k \in \Z_d$, there exists a poly-size normalizer circuit $\mc{C}$ over the group $\Z_d$ that transforms $Z(j) X(k)$ into a diagonal Pauli operator of the form $\gamma^{a} Z(\gcd(j,k))$. Furthermore, there are efficient classical algorithms to compute a description of $\mc{C}$.
\end{lemma}
\begin{corollary}\label{PEG lemma for Abelian groups}
Consider a Pauli operator $\sigma$ over an arbitrary finite Abelian group $G$. Then there exists a poly-size normalizer circuit $\mc{C}$ over $G$ such that ${\cal C}\sigma{\cal C}^{\dagger} =\gamma^{a}Z(g)$. Furthermore, there are efficient classical algorithms to compute a description of $\mc{C}$ as well as  $\gamma^a$, a and $g$.
\end{corollary}
\begin{proof}
To compute $\mc{C}$ note that every Pauli operator over $G$ has the form $\sigma \propto U_1\otimes\cdots\otimes U_m$ where $U_i$ is a Pauli operator over $\mathbb{Z}_{d_i}$ and apply lemma \ref{lem_Nielsen02UniversalSimulations} to each factor. The rest follows by applying theorem \ref{thm_G_circuit_fundamental} to compute the label of $\mc{C}\sigma{\cal C}^{\dagger}$ and, in the case of $\gamma^{a}$, by using standard algorithms to compute scalar exponentials.
\end{proof}
Lemma \ref{lem_Nielsen02UniversalSimulations} and corollary \ref{PEG lemma for Abelian groups} reduce the problem of measuring general Pauli operators to that of implementing measurements of $Z(g)$. Indeed,  given an arbitrary $\sigma$ to be measured, we can always compute a poly-size normalizer circuit that transforms it into a diagonal operator $\gamma^{a}Z(g)$, using corollary \ref{PEG lemma for Abelian groups}.  Then, the measurement of $\sigma$ is equivalent to the procedure (a) apply ${\cal C}$; (b) measure $\gamma^{a}Z(g)$; (c) apply ${\cal C}^{\dagger}$. Finally, Pauli operators that are proportional to each other define the \textit{same} quantum measurement, up to a simple relabeling of the outcomes. Therefore it suffices to focus on the problem of measuring an operator of the form $Z(g)$.

Note now that, by definition, the eigenvalues of $Z(g)$ have the form $\chi_{g}(h)$. Define the following function $\omega$ from $G$ to $\Z_d$, where  $d=\textnormal{lcm}(d_1,\ldots,d_m)$:

\begin{equation}\label{Pauli Measurement - Labelling Function}
\omega(h)= \sum_{i}\, \frac{d}{d_i}\, g(i) h(i) \mod d.
\end{equation}
With this definition one has $\chi_{g}(h) = \euler^{2\pii \omega(h)/d}$. Given any $y\in\mathbb{Z}_d$, the eigenspace of $Z(g)$ belonging to the eigenvalue $\lambda= \euler^{2\pii y/d}$ is spanned by all standard basis states $|h\rangle$ with $\omega(h)=y$.

Note that $\omega$ is a group homomorphism from $G$ to $\Z_d$ as it fulfills (\ref{Homomorphism condition}). As a result, the controlled operation $f(h,a)= (h,a+\omega(h))$ is a group automorphism of $G\times \Z_d$ and it can be implemented by a normalizer gate $U_f\ket{h,a}=\ket{h,a+\omega(h)}$.

The gate  $U_f$ can now be used to measure $Z(g)$, with a routine inspired by the coset-state preparation method used in the standard quantum algorithm to solve the Abelian hidden subgroup problem \cite{lomont_HSP_review, childs_lecture_8}: first, add an auxiliary  $d$-dimensional system $\C^d$ in the state $\ket{0}$ to $\C^{G}$, being the latter in some arbitrary state $\ket{\psi}$; second, apply the global interaction $U_f$; third, measure the ancilla in the standard basis. The global evolution of the system along this process is
\begin{equation*}\label{implementing Pauli measurement Equation}
\ket{\psi}\ket{0}=\sum_{h\in G}\psi(h)\ket{h}\ket{0}\xrightarrow{U_f}\sum_{h\in G}\psi(h)\ket{h}\ket{\omega(h)} \xrightarrow{\textnormal{Measure} } \: \frac{1}{\sqrt{p_y}}\: \left(\sum_{h: \omega(h)=y}\psi(h)\ket{h}\ket{y}\right)
\end{equation*}
The measurement yields an outcome $y\in\mathbb{Z}_d$ with probability $p_y=\sum_{h: \omega(h)=y}|\psi(h)|^2$. The latter precisely coincides with $\|P_{\lambda}\ket{\psi}\|^{2}$ where $P_\lambda$ is the eigenprojector associated with the eigenvalue $\lambda = \euler^{2\pii y/d}$ and, therefore, we have implemented the desired measurement.

\begin{figure}[htbp]
\begin{equation*}
 \Qcircuit @C=1.2em @R=.5em {
  \lstick{\ket{g(1)}}    & \qw & \multigate{3}{\mc{C}}&  \qw & \ctrl{4} & \qw        & \qw & \cdots & & \qw                & \multigate{3}{\mc{C}^{\dagger}} & \qw    & \qw  \\
  \lstick{\ket{g(2)}}    & \qw & \ghost{\mc{C}}  &  \qw & \qw      & \ctrl{3}   & \qw & \cdots & & \qw                & \ghost{\mc{C}^{\dagger}}     & \qw & \qw   \\
  \lstick{\vdots \quad } &    &   \pureghost{\mc{C}}&   &          &            &     &        & &                    & \pureghost{\mc{C}^{\dagger}}    & \vdots &  &     \\
  \lstick{\ket{g(m)}}    & \qw & \ghost{\mc{C}}  & \qw & \qw      & \qw        & \qw & \cdots & & \ctrl{1}           & \ghost{\mc{C}^{\dagger}}     & \qw  &\qw\\	
  \lstick{\ket{0}_d} & \qw &\qw  & \qw & \gate{X_d^{c_1}} & \gate{X_d^{c_2}} & \qw & \cdots & & \gate{X_d^{c_m}} & \meter & \cw & \cw
 }
\end{equation*}
\caption{Quantum circuit implementing measurement of operator $\sigma=\mathcal{C} Z(g) \mathcal{C}^{\dagger}$ }
\label{figure Quantum circuit for Pauli measurements}
\end{figure}
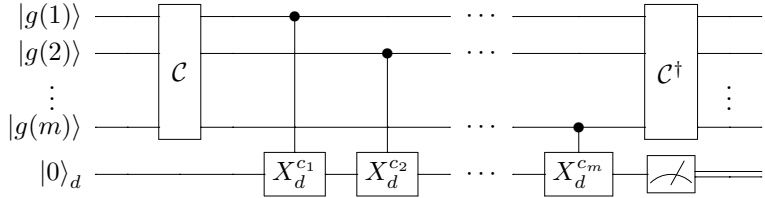

In figure \ref{figure Quantum circuit for Pauli measurements} we show a poly-size quantum circuit that implements the measurement of the Pauli operator $\sigma=\mc{C}Z(g)\mc{C}^{\dagger}$ in the way just described. In the picture, the $m+1$ horizontal lines represent the $m$ physical subsystems that form $\C^{G}={\C^{d_1}\otimes\cdots\otimes\C^{d_m}}$ and the ancillary system ${\C^{d}}$; the numbers $c_i:=d/d_i\, g(i)$ are chosen to compute the function (\ref{Pauli Measurement - Labelling Function}) in the ancillary system. For merely pictorial reasons, the depicted measurement acts on a standard-basis state.

Two  remarks. First, the state of the ancilla could be reset (with Pauli gates) to its original value once the measurement outcome $\omega(x)$ is recorded; this could be used to implement a series of measurements using only one ancilla. Second, the value $\omega(x)$ can be used to compute $\lambda=\chi_{g}(x)=\euler^{2\pii \omega(x)/d }$.

Finally we mention that a procedure given in \cite{deBeaudrap12_linearised_stabiliser_formalism} to implement measurements of qudit Pauli operators (as presented in section \ref{Examples Paulis/Normalizer for qudits}) can be recovered from ours by choosing $G=\Z_d^{m}$.

\subsection{Measurement update rules}

In this section we show that  Pauli measurements transform stabilizer states into new stabilizer states. We give an analytic formula to update their description. Moreover we show that the update can be carried out efficiently.
\begin{theorem}\label{thm_Measurement_Update_rules}
Consider a stabilizer state $\ket{\phi}$ over $G$ with stabilizer group $\mc{S}$ and let $\sigma$ be a Pauli operator. Perform a measurement of $\sigma$ on $|\phi\rangle$, let the measurement outcome be labeled by an eigenvalue $\lambda$ of $\sigma$, and let $|\phi_m\rangle$ denote the post-measurement state.  Then the following statements hold:
\begin{itemize}
\item[(i)] The state $\ket{\phi_m}$  is a stabilizer state, with stabilizer group
\begin{equation}
\mc{S}_m=\langle\, \overline{\lambda}\sigma,\, C_{\mc{S}}(\sigma)\rangle.
\end{equation}
Here $C_{\mc{S}}(\sigma)$ denotes the centralizer of $\sigma$ inside $\mc{S}$, i.e.\ the group containing all elements of $\mc{\mc{S}}$ that commute with $\sigma$.
\item[(ii)] The labels of a generating set of ${\cal S}_m$ can be computed efficiently classically, given the labels of a generating set of ${\cal S}$.
\end{itemize}
\end{theorem}
\begin{proof}
First we show that $\ket{\phi_m}$ is stabilized by $\mc{S}_m$. To see this, first note that $\ket{\phi_m}$ is trivially stabilized by $\overline{\lambda}\sigma$. Furthermore, $\sigma$ commutes with every $\tau\in C_{\mc{S}}(\sigma)$. It follows that the projector $P$ onto the $\lambda$-eigenspace of $\sigma$ commutes with $\tau$ as well (this can easily be shown by considering $\sigma$ and $\tau$ in their joint eigenbasis). Hence $\tau P|\phi\rangle = P\tau |\phi\rangle = P|\phi\rangle$. Using that $\ket{\phi_m}\propto P|\phi\rangle$, we find that $\tau|\phi_m\rangle = |\phi_m\rangle$ for every $\tau\in C_{\mc{S}}(\sigma)$.

Second, we prove that $\ket{\phi_m}$ is the unique state stabilized by the group $\mc{S}_m$.  Without loss of generality we  restrict to the case $\sigma=Z(g_m)$. This is sufficient since, first, every Pauli operator can be transformed into an operator of the form $\alpha Z(g)$ with a suitable normalizer circuit $\mathcal{C}$ (cf. the discussion in section \ref{section  Pauli measurements Implementation}); and, second,  for any normalizer circuit $\mc{C}$, the quantum state $\ket{\phi_m}$ is a stabilizer state with stabilizer group $\mc{S}_m$ if and only if $\mc{C}\ket{\psi_m}$ is a stabilizer state with stabilizer group $\mc{C}\mc{S}_m\mc{C}^{\dagger}$.

Working with the assumption $\sigma=Z(g_m)$, we write the label subgroups $\mathbb{H}_{m}$ and $\mathbb{D}_{m}$ of $\mc{S}_m$ in terms of the label groups  $\mathbb{H}$ and  $\mathbb{D}$ of ${\cal S}$.  We have $\mc{S}_m=\langle\, \overline{\lambda}\sigma,\, C_{\mc{S}}(\sigma)\,\rangle$ where $\sigma=Z(g_m)$ for some $g_m\in G$. This implies that only the labels of  $C_{\mc{S}}(\sigma)$ contribute to $\mathbb{H}_{m}$. The centralizer $C_{\mc{S}}(\sigma)$ can be written as $C_{\mc{S}}(\sigma)=\mc{S}\cap C_{\mc{P}}(\sigma)$,  where $C_{\mc{P}}(\sigma)$ is  the subgroup of all Pauli operators that commute with $\sigma$. Thence, using lemma \ref{Commutativity of Pauli Stabilizer Groups PROPOSITION} we see that $C_{\mc{P}}(\sigma)$ consists of all $\gamma^{a}Z(g)X(h)$ with labels $h\in \langle g_m\rangle^{\perp}$. Hence,
\begin{equation}\label{H_m}
\mathbb{H}_{m}=\mathbb{H}\cap \langle g_m\rangle^{\perp}
\end{equation}
Due to the commutativity of $Z(g_m)$ and $C_{\mc{S}}(\sigma)$,  any element in $\mc{S}_m$  can be reordered as $\tau\, Z(g_m)^{i}$, with $\tau\in C_{\mc{S}}(\sigma)$.  Therefore, the diagonal group of ${\mc{S}_m}$ can be written as $\mc{D}_{m}= \langle\,\mc{D}',\, Z(g_m) \rangle $ where ${\cal D}'$ is the diagonal subgroup of $C_{\mc{S}}(\sigma)$. We now claim that ${\cal D}' ={\cal D}$ where ${\cal D}$ is the diagonal subgroup of ${\cal S}$. To see this, first note that trivially ${\cal D}' \subseteq {\cal D}$ since $C_{\mc{S}}(\sigma)$ is a subgroup of ${\cal S}$. Conversely, ${\cal D} \subseteq {\cal D}'$: as every diagonal element of ${\cal S}$ commutes with $Z(g_m)$, we have ${\cal D}\subseteq C_{\mc{S}}(\sigma)$; but this implies ${\cal D} \subseteq {\cal D}'$.

Putting everything together, we thus find $\mc{D}_{m}= \langle\,\mc{D},\, Z(g_m) \rangle $. It follows:
\begin{equation}\label{D_mH_m}
\mathbb{D}_{m}=\langle \mathbb{D}, \langle g_m\rangle \rangle \quad\Longrightarrow\quad \mathbb{D}_{m}^{\perp}=\left\langle \mathbb{D}, \langle g_m\rangle\right\rangle^{\perp}=\mathbb{D}^{\perp}\cap \langle g_m\rangle^{\perp},
\end{equation}
where we used lemma \ref{properties of orthogonal subgroups}. Since ${\cal S}$ uniquely stabilizes $|\phi\rangle$, we have $\mathbb{H}=\mathbb{D}^{\perp}$ owing to corollary \ref{corollary_Uniqueness_Test}. With (\ref{D_mH_m}) and (\ref{H_m}) this implies that $\mathbb{H}_{m}=\mathbb{D}_{m}^{\perp}$. Again using corollary \ref{corollary_Uniqueness_Test}, it follows that ${\cal S}_m$ uniquely stabilizes $|\phi_m\rangle$.\\

To complete the proof of the theorem, we give an efficient classical algorithm to find a generating set of the centralizer $C_{\mc{S}}(\sigma)$; our approach is to reduce this task to a certain problem over the group $G\times G$ that can be efficiently solved using lemma \ref{lemma Algorithms for finite Abelian groups}. Let $K\subset G\times G$  be the group of tuples $(g,h)$ such that there exists a stabilizer operator $\sigma(a,g,h)\in C_{\mc{S}}(\sigma)$; we prove that  $C_{\mc{S}}(\sigma)$ is isomorphic to $K$ via the map $\kappa : \sigma(a,g,h) \rightarrow (g,h)$, and that $\kappa$ is efficiently classically invertible: this  reduces the problem to finding a generating set of $K$ and applying the map $\kappa^{-1}$ to all its elements.

First, it is straightforward to verify that $\kappa$ is an isomorphism. Equations (\ref{commutation relations Pauli operators}) imply that the map is indeed linear. Surjectivity is granted by definition. Invertibility follows then from the fact that only  elements of the type $\gamma^{a}I\in C_\mathcal{S}(\sigma)$, for some $a$,  belong to $\ker\kappa$ (where $I$ denotes the identity): the latter are invalid stabilizer operators unless $\gamma^a=1$.

Second, we show how to compute $\kappa^{-1}$. Let the operator to measure $\sigma$ be of the (general) form  $\sigma=\gamma^{a_m}Z(x)X(y)$, and let $(a',g',h')$ be the label of an arbitrary stabilizer $\tau\in \mathcal{S}$. Given a set of (mutually commuting) generators $\sigma_1, \ldots, \sigma_r$ of $\mc{S}$, with corresponding labels $(a_i,g_i,h_i)$, the element $\tau$ can be written in terms of them as \be \label{group isomorphic to centralizer1}\tau=\prod \sigma_{i}^{v_i}= \gamma^{a'} X\left(\sum v_i g_i \right) Z\left(\sum v_i h_i \right)\ee for some integers $v_i$. From this equation it follows that
$K\subset \langle (g_1, h_1),\ldots,(g_r, h_r) \rangle$, which leads us to the following algorithm to compute $\kappa^{-1}$: given $(g,h)\in K$,  use the algorithm of lemma \ref{lemma Algorithms for finite Abelian groups}(a) to compute $r$ integers $w_i$ such that $({g},h)=\sum w_i(g_i,h_i)$; due to (\ref{commutation relations Pauli operators}), the stabilizer operator defined as $\varsigma=\prod \sigma_i^{w_i}$ (whose label can be efficiently computed) is proportional to $X(g)Z({h})$; it follows that $\kappa(\varsigma)=(g,h)$ and, hence, $\varsigma$ equals $\kappa^{-1}(g,h)$.

Finally, combining (\ref{group isomorphic to centralizer1}) with formula (iii) in lemma  \ref{Commutativity of Pauli Stabilizer Groups PROPOSITION} we obtain
\begin{equation}\label{group isomorphic to centralizer2}
K=\langle y,-x\rangle^{\perp}\cap \langle (g_1, h_1),\ldots,(g_r, h_r) \rangle.
\end{equation}
Using eq.\ (\ref{group isomorphic to centralizer2}) together with algorithms (c-d) of lemma \ref{lemma Algorithms for finite Abelian groups}, we can efficiently compute $s=\ppolylog{\Inputsize}$ elements $(x_1,y_1),\ldots,(x_s,y_s)$ that generate $K$; applying $\kappa^{-1}$ to these, we end up with a set of stabilizer operators $\kappa^{-1}(x_i,y_i)$ that generates $C_{\mc{S}}(\sigma)$.
\end{proof}

\section{Classical Simulation of Adaptive Normalizer Circuits \label{section Gottesman-Knill theorem}}

Recall that in \cite{VDNest_12_QFTs} the following classical simulation result was shown:
\begin{theorem}\label{thm_maarten}
Let  $G=\mathbb{ Z}_{d_1}\times\cdots \times \mathbb{ Z}_{d_m}$ be a finite Abelian group. Consider a polynomial size unitary normalizer circuit over $G$ acting on a standard basis input state. Both circuit and input are specified in terms of their standard encodings as described above. The circuit is followed by a measurement in the standard basis. Then there exists an efficient classical algorithm to sample the corresponding output distribution.
\end{theorem}
In the theorem, the \textit{standard encoding} of a normalizer circuit is defined as in section \ref{section Pauli and Clifford and Normalizer} in this work; the \textit{standard encoding} of a standard basis input state $\ket{g}$ is simply the tuple $g$, i.e. a collection of $m$ integers. Recall also that ``efficient'' is synonymous to ``in polynomial time in $\log |G|$''.

The main classical simulation result of this paper (theorem \ref{thm_main} below) is a generalization of the above result. Rather than unitary normalizer circuits, the family of quantum circuits  considered here is that of the \emph{adaptive normalizer circuits}. A polynomial-size adaptive normalizer circuit consists of \ppolylog{\Inputsize} elementary steps, each of which is either a unitary normalizer gate $U$ or a Pauli measurement $M$. Furthermore, the choice of which $U$ or $M$ to apply in any given step may depend, in a (classical) polynomial-time computable way, on the collection of outcomes obtained in all previous measurements. The notion of adaptive normalizer circuits is thus a direct generalization of the adaptive Clifford circuits considered in the original Gottesman-Knill theorem \cite{Gottesman_PhD_Thesis, gottesman_knill}. Note that, compared to theorem \ref{thm_maarten}, two elements are added. First, measurements are no longer restricted to be standard basis measurements but arbitrary Pauli measurements. Second, the circuits are adaptive.

Before stating our classical simulation result, we make precise what is meant by an efficient classical simulation of an adaptive normalizer circuit. First, recall that the outcomes of any Pauli measurement are labeled by the eigenvalues of the associated Pauli operator. Since $\sigma^{2\mathfrak{g}}=I$ (recall lemma \ref{pauli_products_powers PROPERTY}) it follows that each Pauli operator eigenvalue is a $2\mathfrak{g}$-th root of unity i.e.\ it has the form $\lambda = \euler^{\pii k/\mathfrak{g}}$ for some $k\in \{0,\ldots, 2\mathfrak{g}-1\}$. This implies that any Pauli measurement gives rise to a probability distribution over the set of $2\mathfrak{g}$-th roots of unity; we denote the latter set by $S_{2\mathfrak{g}}$. Now consider an adaptive normalizer circuit ${\cal C}$. Let $P_i(\lambda | \lambda_1 \cdots \lambda_{i-1})$ denote the conditional probability of obtaining the outcome $\lambda\in S_{2\mathfrak{g}}$ in the $i$-th measurement, given that in previous measurements the outcomes $\lambda_1 \cdots \lambda_{i-1}\in S_{2\mathfrak{g}}$ were measured. We now say that ${\cal C}$ can be simulated efficiently classically if \textit{for every $i$} the $i$-th conditional probability distribution $P_i(\lambda | \lambda_1 \cdots \lambda_{i-1})$ can be sampled efficiently on a classical computer,  given the description of all gates and measurement operators in the circuit.
\begin{theorem}[\textbf{Classical simulation of adaptive normalizer circuits}]\label{thm_main}
Consider a polynomial size adaptive normalizer circuit over $G$, specified in terms of its standard encoding, which  acts on an arbitrary standard basis input state. Then any such circuit can be efficiently simulated classically.
\end{theorem}
\begin{proof}
Let ${\cal C}$ denote the adaptive normalizer circuit. Without loss of generality we assume that the input state is $|0\rangle$. Indeed, any standard basis state $|g\rangle$ can be written as $|g\rangle = X(g)|0\rangle$; the Pauli operator $X(g)$ can be realized as a polynomial-size normalizer circuit \cite{VDNest_12_QFTs} and can thus be absorbed in the overall adaptive normalizer circuit. Letting $e_i\in G$ denote the $i$-th ``canonical basis vector'', the state $|0\rangle$ is a stabilizer state with stabilizer generators $Z(e_1),\ldots, Z(e_m)$ \cite{VDNest_12_QFTs}. We now recall the following facts, proved above:
\begin{itemize}
\item[(a)] Given any normalizer gate $U$ and any stabilizer state $|\psi\rangle $ specified in terms of a generating set of \ppolylog{\Inputsize} generators, the state $U|\psi\rangle$ is again a stabilizer state; moreover a set of generators can be determined efficiently (see theorem \ref{thm_G_circuit_fundamental}).
\item[(b)] Given any Pauli operator $\sigma$ and any stabilizer state $|\psi\rangle $ specified in terms of a generating set of polylog$(\mathfrak{g})$ generators, the state $|\psi_{\lambda}\rangle$ obtained after measurement of $\sigma$, for any outcome $\lambda$, is again a stabilizer state; moreover a set of generators can be determined efficiently (cf. theorem \ref{thm_Measurement_Update_rules}).

    Furthermore, the measurement probability distribution can be sampled efficiently in polynomial time on a classical computer. The latter is argued as follows. First, it follows from the discussion in section \ref{section  Pauli measurements Implementation} that the simulation of any Pauli measurement, on some input stabilizer state $\ket{\psi}$, reduces to simulating a unitary normalizer circuit (the description of which can be computed efficiently) followed by a \textit{standard basis} measurements (acting on the same input $\ket{\psi}$ and a suitable ancillary stabilizer state $\ket{0}$). Second, normalizer circuits acting on stabilizer state inputs and followed by standard basis measurements on stabilizer states can be simulated efficiently: this was proved for the special case of coset state inputs in \cite{VDNest_12_QFTs}; their argument, however, carries over immediately to the general case of stabilizer state inputs.
\end{itemize}
The proof of the result is now straightforward. Given any tuple $\lambda_1,\ldots, \lambda_{i-1}$, a generating set of stabilizers can be computed efficiently for the state of the quantum register obtained immediately before the $i$-th measurement, given that the previous measurement outcomes were $\lambda_1 \cdots \lambda_{i-1}$. Furthermore, given this stabilizer description, the distribution $P_i(\lambda | \lambda_1 \cdots \lambda_{i-1})$ can be sampled efficiently on a classical computer, as argued in (b).
\end{proof}
To conclude this section we comment on an interesting difference between normalizer circuits and the ``standard'' qubit Clifford circuits, concerning the role of adaptiveness as a \textbf{tool for state preparation}. For qubits, adaptiveness adds no new state preparation power to the unitary Clifford operations. Indeed for any $n$-qubit stabilizer state $\ket{\psi}$ there exists a (poly-size) \textit{unitary} Clifford circuit $\mathcal{C}$ such that $\ket{\psi}=\gamma\mathcal{C}\ket{0}^{n}$, for some global phase $\gamma$ \cite{dehaene_demoor_coefficients}. In contrast, over general Abelian groups $G$ this feature is no longer true. The associated adaptive normalizer circuits allow to prepare a \textit{strictly} larger class of stabilizer states compared to unitary normalizer circuits alone.

To demonstrate this claim, we provide a simple example of a stabilizer state over $G=\Z_{4}$  that \textit{cannot} be prepared from standard basis input states via unitary normalizer transformations over $G$, even in exponential time. However, the same state can be prepared efficiently  \textit{deterministically} if one considers adaptive normalizer schemes.
We consider
\begin{equation}\label{Peculiar stabilizer state}
\ket{{\psi}}=\frac{1}{\sqrt{2}}\left(\ket{0}+\ket{2}\right)
\end{equation}
Suppose that there existed a unitary Clifford operator $U\in\mc{C}^{G}$ which generates $\ket{{\psi}}$ from $|0\rangle$. Since the stabilizer group of $\ket{0}$ is  generated by $Z_d$, the stabilizer group of $\ket{\psi}$ would be generated by  $U Z_d U^{\dagger}$. However it was shown in \cite{dehaene_demoor_hostens} that the stabilizer group of $\ket{{\psi}}$  cannot be generated by {one} single Pauli operator (i.e. at least two generators are needed), thus leading to a contradiction.

On the other hand, we now provide an efficient adaptive normalizer scheme to prepare,  not only the example $\ket{\psi}$, but in fact any coset state \cite{lomont_HSP_review,childs_lecture_8,childs_vandam_10_qu_algorithms_algebraic_problems}  of any finite Abelian group $G$. This refers to any state of the form
\begin{equation}
|H+x\rangle:=\frac{1}{\sqrt{|H|}}\sum_{h\in H}\ket{h+x},
\end{equation}
where $H$ is a subgroup of $G$ and $x\in G$. Note that $|\psi\rangle$ is a coset state of the group $G=\Z_4$ with $H:=\langle 2\rangle$ and $x:=0$.

Our algorithm to efficiently prepare general coset states $|H+x\rangle$ receives the element $x$ and a polynomial number of generators of $H$. In section \ref{sect_CompComp_FiniteAbelianGroups} (in the example after lemma \ref{LEMMA REDUCTION TO SYSTEMS OF LINEAR EQUATIONS}) we showed how to efficiently compute the matrix representation of a group homomorphism $\varpi:G\rightarrow \Z_d^{s}$ such that $\ker{\varpi}=H$, where the integer $s$ is $O(\polylog{\Inputsize})$ and $d=\Inputsize$. Given $\varpi$, we define a group automorphism $\alpha$ of the group $G\times \Z_d^{s}$ by $\alpha(g,h):=(g,h+\varpi(g))$. We now consider the following procedure\footnote{Observe that $\varpi$ can be considered as a function that \textit{hides} the subgroup $H$ in the sense of the hidden subgroup problem (HSP) \cite{lomont_HSP_review,childs_lecture_8,childs_vandam_10_qu_algorithms_algebraic_problems}. That is, for every $g, g' \in G$ we have $\varpi(g)=\varpi(g')$ iff $g-g'\in H$. Procedure (\ref{coset_preparation}) is essentially the routine used in the quantum algorithm for HSP to prepare random coset states.}:
\begin{equation}\label{coset_preparation}
\ket{0}\ket{0} \xrightarrow{F\otimes I} \sum_{h\in G} \ket{h}\ket{0} \xrightarrow{U_{\alpha}} \sum_{h\in G} \ket{h}\ket{\varpi(h)} \xrightarrow{{M}} \: \frac{1}{\sqrt{|H|}}\: \sum_{h\in H}\ket{g+h}\ket{b} = \ket{g+H}\ket{b},
\end{equation}
where $F$ denotes the QFT over $G$, the unitary $U_{\alpha}$ is the automorphism gate sending $\ket{g,h}$ to $\ket{\alpha(g,h)}$, and $M$ is a measurement of the second register in the standard basis. If the measurement outcome is $b$, then the post-measurement state is $\ket{g+H}\ket{b}$ where $g$ is a solution of the equation $\varpi(g)= b$. It can be verified that each coset state of $H$ (and thus also the desired coset state $\ket{x+H}$) occurs equally likely, i.e. with  probability $p=|H|/|G|$: in general, $p$ can be exponentially small. However, if we apply adaptive operations, we can always prepare $\ket{x+H}$ with probability 1, as follows. First, given the measurement outcome $b$ we efficiently compute an element $g'\in G$ satisfying  $\varpi(g')=b$ using theorem \ref{THM SYSTEM OF LINEAR CONGRUENCES MODULO AN ABELIAN GROUP}. Then we apply a ``correcting'' Pauli operation $X(x-g')$ to the first register state, yielding $X(x-g')\ket{g+H}=\ket{x +(g-g')+H}=\ket{x+H}$ as desired (we implicitly used $g-g'\in H$).

\section{Normal Form of a Stabilizer State \label{section Normal form of stabilizer states}}

 In this section we give an analytic characterization of the amplitudes of arbitrary stabilizer states over finite Abelian groups. In addition, we show that these amplitudes can be efficiently classically computed.
\begin{theorem}[\textbf{Normal form of an stabilizer state}]\label{thm Normal form of an stabilizer state}
Every stabilizer state $\ket{\phi}$ over a finite Abelian group $G$ with stabilizer group $\mc{S}$  has the form
\begin{equation}\label{Normal form of an stabilizer state EQUATION}
\ket{\phi}=\alpha \frac{1}{\sqrt{|\mathbb{H}|}}\sum_{h\in \mathbb{H}}\xi(h)\ket{s+h}.
\end{equation}
Here $\alpha$ is a global phase, $\mathbb{H}$ is the label group (\ref{Label group H}), $s\in G$, and relative phases are described by a quadratic function $\xi$ on the group $\mathbb{H}$.  Furthermore, if a generating set $\{\sigma_1,\ldots,\sigma_r\}$ of $\mc{S}$ is specified, the following tasks can be carried out efficiently:

(a) Compute $s$;

(b) Given $g\in G$, determine if $g\in s + \mathbb{H}$;

(c) Given $h\in \mathbb{H}$, compute $\xi(h)$ up to $n$ bits in $\ppoly{n, \log \mathfrak{g}}$ time;

(d) Compute $\sqrt{|\mathbb{H}|}$.
\end{theorem}
\begin{proof}
Corollary \ref{corollary_Uniqueness_Test} implies that $\mathbb{D}^{\perp} = \mathbb{H}$. Using this identity together with theorem \ref{thm structure test}(i), we find that supp($|\phi\rangle) = s+ \mathbb{H}$ for some $s\in G$. By definition of $\mathbb{H}$, for every $h\in\mathbb{H}$ there exists some element $\sigma(a,g,h)\in\mc{S}$. Using that $\sigma(a,g,h)|\phi\rangle = |\phi\rangle$ we then have
\begin{equation}\label{Quadratic Relative Phases DEFINITION}
\langle s + h | {\phi} \rangle=\langle s + h |\sigma(a,g,h)|{\phi} \rangle= {\gamma^{a}\chi_{s+h}(g)}\langle s |{\phi} \rangle
\end{equation}
This implies that $|\langle s + h | {\phi} \rangle| = |\langle s| {\phi} \rangle|$ for all $h\in \mathbb{H}$. Together with the property that supp($|\phi\rangle) = s+ \mathbb{H}$, it follows that $|\phi\rangle$ can be written as
\begin{equation}\label{uniform supperposition}
|\phi\rangle = \frac{1}{\sqrt{|\mathbb{H}|}} \sum_{h\in\mathbb{H}} \xi(h) |s+h\rangle
\end{equation}  for some complex phases $\xi(h)$. By suitably choosing an (irrelevant) global phase, w.l.o.g. we can assume that $\xi(0)=1$.

We now show that the function $ h\in H \to \xi(h)$ is  quadratic. Using (\ref{Quadratic Relative Phases DEFINITION}, \ref{uniform supperposition}) we derive \be\label{proof_relphases_are_quadratic0}\xi(h) = \sqrt{|\mathbb{H}|} \langle s+h|\phi\rangle = \sqrt{|\mathbb{H}|}{\gamma^{a}\chi_{s+h}(g)}\langle s |{\phi} \rangle = \gamma^{a}\chi_{s+h}(g)\xi(0)=\gamma^{a}\chi_{s+h}(g).\ee Since $\xi(h)$ by definition only depends on $h$, the quantity $\gamma^{a}\chi_{s+h}(g)$ only depends on $h$ as well: i.e.\ it is independent of $a$ and $g$. Now select $h_1, h_2\in \mathbb{H}$ and the associated stabilizer operators $\sigma_1(a_1,g_1,h_1)$, $\sigma_2(a_2,g_2,h_2)\in {\cal S}$. Then
\begin{align}
\xi(h_1 + h_2)&= \sqrt{|\mathbb{H}|}\langle s + h_1+h_2 | {\phi} \rangle\\
&= \sqrt{|\mathbb{H}|}\langle s + h_1+h_2 | \sigma_1 \sigma_2 | {\phi} \rangle \label{proof_relphases_are_quadratic1}\\
&= \sqrt{|\mathbb{H}|} \gamma^{a_1}\chi_{s+h_1+h_2}(g_1)\: \langle s + h_2 |  \sigma_2 | {\phi} \rangle \label{proof_relphases_are_quadratic2}\\
&= \left[ \gamma^{a_1}\chi_{s+h_1}(g_1) \right] \: \left[ \gamma^{a_2}\chi_{s+h_2}(g_2)\right]  \: \chi_{g_1}(h_2) \: \xi(0) \label{proof_relphases_are_quadratic3} \\ &= \xi(h_1) \xi(h_2) \chi_{g_1}(h_2)  \label{proof_relphases_are_quadratic4}
\end{align}
In (\ref{proof_relphases_are_quadratic1}) we used that $\sigma_1\sigma_2|\phi\rangle = |\phi\rangle$; in (\ref{proof_relphases_are_quadratic2}-\ref{proof_relphases_are_quadratic3}) we used the definitions of Pauli operators and the fact that $\langle s|\phi\rangle = \xi(0)$; finally in (\ref{proof_relphases_are_quadratic4}) we used identity (\ref{proof_relphases_are_quadratic0}) and the fact that $\xi(0)=1$.  Now define $B(h_1,h_2)= \xi(h_1+h_2)\overline \xi(h_1) \overline \xi(h_2)$. We  claim that $B$ is a bilinear function of $\mathbb{H}$.  To see this, note that the derivation above shows that $B(h_1, h_2)=\chi_{g_1}(h_2)$. Linearity in the second argument $h_2$ is immediate. Furthermore, by definition $B$ is a symmetric function i.e.\ $B(h_1,h_2) = B(h_2,h_1)$.  This shows that $B$ is bilinear, as desired.

We now address (a)-(d). As for (a) recall that  $s+\mathbb{H}$ is the support of a stabilizer state  $|\phi\rangle$; theorem \ref{thm structure test} then provides an efficient method to compute a suitable representative $s$. Note also that a generating set of $\mathbb{H}$ can be computed efficiently owing to lemma \ref{label groups of an stabilizer DEFINITION}. Statement (b) follows from lemma \ref{lemma Algorithms for finite Abelian groups}(a). Statement (d) follows from lemma \ref{lemma Algorithms for finite Abelian groups}(b). Finally we prove (c), by showing that the following procedure to compute $\xi(h)$ is efficient, given any $h\in \mathbb{H}$:

\vspace{2mm}

(i) determine some element $\sigma \in {\cal S}$ such that $\sigma|s\rangle \propto |s+h\rangle$;

(ii) compute $\langle s+h|\sigma|s\rangle = \xi(h)$.

\vspace{2mm}

\noindent To achieve (i), it suffices to determine an arbitrary stabilizer element of the form $\sigma = \sigma(a,g,h)\in\mc{S}$. Assume that generators $\sigma_1(a_1,g_1,h_1),\ldots,\sigma_r(a_r,g_r,h_r)$ are given to us. We can then use algorithm (a) in lemma \ref{lemma Algorithms for finite Abelian groups} to find integers $w_i$ such that $h=\sum w_i h_i$ and, due to, $\sigma = \prod \sigma_i^{w_i}$ is an operator of form $\sigma(a,g,h)$ for some values of $a,g$ (use eq.\ \ref{commutation relations Pauli operators}). Moreover, given the $w_i$ the label $(a,g,h)$ of $\sigma$ can be computed efficiently; this accomplishes (i). Finally, it is straightforward that (ii) can be carried out efficiently: using formula $\xi(h)=\gamma^{a}\chi_{s+h}(g)$ and standard algorithms to compute elementary functions \cite{brent_zimmerman10CompArithmetic}.
\end{proof}
Theorem \ref{thm Normal form of an stabilizer state} generalizes result from \cite{dehaene_demoor_coefficients,dehaene_demoor_hostens} where analogous characterizations were given for qubits and qudits, although those works do not consider the notion of quadratic functions used here (furthermore their methods are completely different from ours). For example, in ref.\ \cite{dehaene_demoor_coefficients} it was shown that every Pauli stabilizer state for qubits (corresponding to the group $\mathbb{Z}_2^m$) can be written as \be |\phi\rangle \propto \frac{1}{\sqrt{|S|}} \sum_{x\in S} (-1)^{q(x)} i^{l(x)} |x+s\rangle.\ee Here $S$ is a linear subspace of $\mathbb{Z}_2^m$, $q(x) = x^T Ax\mod{2}$ is a quadratic form over $\mathbb{Z}_2$, and $l(x)\mod{2}$ is a linear form. This characterization indeed conforms with theorem \ref{thm Normal form of an stabilizer state}: the set $S$ is a subgroup of $\mathbb{Z}_2^m$ and the function \be x\in \mathbb{Z}_2^m \to \xi(x):= (-1)^{q(x)} i^{l(x)}\ee is quadratic (see section \ref{section Quadratic functions}).

Theorem \ref{thm Normal form of an stabilizer state} also implies that every stabilizer state belongs to the family of Computationally Tractable states (CT states). A state $|\psi\rangle = \sum \psi_g|g\rangle \in \C^G$ is said to be CT (relative to its classical description) if the following properties are satisfied:
\begin{itemize}
\item[(a)] there exists an efficient randomized classical algorithm to sample the distribution $\{|\psi_g|^2\}$;
\item[(b)] given $g\in G$, the coefficient $\psi_g$ can be computed efficiently with exponential precision.
\end{itemize}
CT states form a basic component in a general class of quantum computations that can be simulated efficiently classically using probabilistic simulation methods. For example consider a quantum circuit ${\cal C}$ acting on a CT state and followed by a final standard basis measurement on one of the qubits. Then, regardless of which CT state is considered, such computation can be efficiently simulated classically when ${\cal C}$ is e.g.\ an arbitrary Clifford circuit, matchgate circuit, constant-depth circuit or sparse unitary. See \cite{nest_weak_simulations} for an extensive discussion of classical simulations with CT states.

Here we show that every stabilizer state $|\psi\rangle\in \C^G$ over a finite Abelian group $G$ is CT. To be precise, we prove that such states are CT \textit{up to a global phase}. That is, instead of (b) we prove a slightly weaker statement which takes into account the fact that any stabilizer state specified in terms of its stabilizer is only determined up to an overall phase. Formally, we consider the property
\begin{itemize}
\item[(b')] there exists an efficient classical algorithm that, on input of $g\in G$, computes a coefficient $\psi_g'$, where the collection of coefficients $\{\psi_g':g\in G\}$ is such that $|\psi\rangle = \alpha \sum \psi_g'|g\rangle$ for some complex phase $\alpha$.
\end{itemize}
\begin{corollary}
Let $|\psi\rangle$ be a stabilizer state over an Abelian group $G$, specified in terms of a generating set of \ppolylog{\mathfrak{g}} stabilizers. Then $|\psi\rangle$ is CT in the sense (a)-(b').
\end{corollary}
\begin{proof}
Property (a) was proved in \cite{VDNest_12_QFTs}. To prove (b'), note that theorem  \ref{thm Normal form of an stabilizer state} implies there exists a global phase $\alpha$ such that \be \langle g|\psi\rangle = \left\{ \begin{array}{cl}\alpha \cdot \frac{1}{\sqrt{|\mathbb{H}|}}\cdot  \xi(h) & \mbox{ if } g=s+h \mbox{ for some } h\in \mathbb{H} \\ 0 & \mbox{ if } g\notin \mathbb{H}+ s.  \end{array}\right.\ee
Using theorem \ref{thm Normal form of an stabilizer state}(b) it can be efficiently determined whether $g$ belongs to $\mathbb{H}+s$. If not, then $\langle g|\psi\rangle=0$. If yes, then compute $h:g-s$; then $\xi(h)$ can be computed owing to theorem  \ref{thm Normal form of an stabilizer state}(c). Finally, $\sqrt{|\mathbb{H}|}$ can be computed owing to theorem  \ref{thm Normal form of an stabilizer state}(d).
\end{proof}

\section{Acknowledgments}

 We are thankful to Arne Storjohann for pointing us reference \cite{storjohann10_phd_thesis}, and grateful to Earl T. Campbell and to the anonymous referees for suggestions to improve presentation. JBV acknowledges financial support from the Elite Network of Bavaria (ENB) program QCCC.

\newpage

\phantomsection

\bibliographystyle{utphys}

\bibliography{references}
\addcontentsline{toc}{section}{References}

\appendix

\section{Proof of lemma \ref{LEMMA REDUCTION TO SYSTEMS OF LINEAR EQUATIONS}}\label{Appendix: Algorithms for Finite Abelian Groups}

\begin{quote}
``\textit{Problems (a-e) in lemma \ref{lemma Algorithms for finite Abelian groups} are polynomial-time reducible to either counting or finding solutions of systems of equations of the form $\mathcal{A}(x)=b$; where $\mc{A}$ is a group homomorphism between two (canonically-decomposed) finite Abelian groups, $\mbf{G}_{{sol}}$ and $\mbf{G}$, to which $x$, $b$ respectively belong; given that a matrix representation of $\mathcal{A}$ is provided.}''
\end{quote}
In the example given in section \ref{sect_CompComp_FiniteAbelianGroups} we proved the lemma for the (d,e)th cases of lemma \ref{lemma Algorithms for finite Abelian groups}; to prove it for each of the remaining cases, (a-c), we will take  similar steps. In the following, we define $A_H$, $A_K$ to be integer matrices whose columns are the elements of the sets $\{h_1,\ldots,h_r\}$ and $\{k_1,\ldots,k_s\}$; the latter generate respectively $H$ and $K$. Also, we  denote by $d$  the least common multiplier of $d_1,\ldots,d_m$. One can  use condition (\ref{Homomorphism condition}) to check that the matrices $A_H$, $A_K$ and $[A_H|A_K]$ define group homomorphisms from $\Z_d^{t}$ to $G$, if the value of $t$ is respectively chosen to be $r$, $s$ and $r+s$.

We will  show how to turn the problems (a-c) into system  of the form $Ax = b\pmod{\mbf{G}}$ such that $\mbf{G}$ equals the original group $G$;  $\mbf{G}_{sol}$ is chosen to be $\Z_d^{t}$, for some $t$; and $A$ is an integer matrix that defines a group homomorphism from $\mbf{G}$ to $\mbf{G}_{sol}$: \\

\textbf{(a)} $b$ belongs to $H$ if and only if $b$ can be obtained as a linear combination of elements of $H$, i.e.,  if and only if $A_H x = b \pmod G$ has at least one solution $x\in \Z_d^{r}$. Moreover, if one finds a particular solution $w$, this element fulfills $b=A_H w = \sum w(i)h_i \pmod{G}$.\\

\textbf{(b)} The order of $H$ is the number of distinct linear combinations of columns of $A_H$, which coincides with the order of the \textit{image} of the group homomorphism  $A_H:\Z_d^{r}\rightarrow G$. With this knowledge, it suffices to count the number of solutions of $A_H x = 0 \pmod G$, which equals $\ker A_H$. Then, one can compute  $|H|=|\textnormal{im}A_H|=d^r/|\ker A_H|$, where the latter identity comes from the first isomorphism theorem ($\text{im}{A_H}\cong \Z_d^{r}/\ker{A_H}$).\\

\textbf{(c)}  $g$ belongs to  $H\cap K$ iff it can be simultaneously written as $h=\sum x(i) h_i = \sum y(i) k_i$ for some  $(x,y)\in \Z_d^{r}\times\Z_d^{s}$; or, equivalently, iff there exist an element $(x,y)$ of the kernel of $\left[A_H | A_K\right]: \Z_d^{r}\times\Z_d^{s}\rightarrow G$ such that $h=A_H x=-A_Ky\pmod{G}$. Thus, given a generating-set $\{(x_i,y_i)\}$ of  $\ker\left[A_H | A_K\right]$, the elements $g_i:=A_H x_i\pmod{G}$ generate $H\cap K$, and, owing to, the problem reduces to finding  solutions of $\left[A_H | A_K\right]\binom{x}{y}=0\pmod{G}$.\\

Finally, notice that $r$, $s$ and $r+s$ are $O(\polylog{\Inputsize})$ due to the initial assumption that the generating-sets are poly-size, and that $d$ is $O(d_1 d_2\cdots d_m)=O(|\mathbf{G}|)$; as a consequence, $\log|\mbf{G}_{sol}|$, $\log|\mbf{G}|$ are also $O(\polylog{\Inputsize})$; and, thus, we need $O(\polylog{\mathfrak{g}})$ memory to store the matrix $A$. It follows that the input-size of the new problem is $O(\polylog{\mathfrak{g}})$ and, therefore, we have reduced all problems (a-c) to systems of linear equations over finite Abelian groups in polynomial time.

\section{Proof of theorem \ref{THM SYSTEM OF LINEAR CONGRUENCES MODULO AN ABELIAN GROUP}}\label{Appendix: System of linear congruences modulo Abelian group}

 We recall here theorem \ref{THM SYSTEM OF LINEAR CONGRUENCES MODULO AN ABELIAN GROUP}.\\

 {\bf Theorem 1} (\textbf{Systems of linear equations over finite Abelian groups}) {\it Given any element $b$ of the group $G=\DProd{d}{m}$ and any $m\times n$  matrix $A$ which defines a group homomorphism from  $H=\DProd{c}{n}$  to $G$, consider the system of equations $A x = b \pmod{G}$. Then, there exist classical algorithms to solve the following list of problems in \ppolylog{{|H|}, {|G|}} time.}
\begin{enumerate*}
\item {\it Decide whether the system admits a solution.}
\item {\it Count the number of different solutions of the system.}
\item {\it Find $x_0, x_1, \ldots, x_r\in H$ such that all solutions of the system are linear combinations of the form $x_0+\sum k_i x_i$.}
\end{enumerate*}

Given $b\in G=\DProd{d}{m}$ and the $m\times n$  matrix $A$, which defines a group homomorphism from  $H=\DProd{c}{n}$  to $G$, we can see the system of linear equations $A x = b \pmod{G}$ as an ``inhomogeneous system of congruences":
\begin{equation}\label{INITIAL SYSTEM OF LINEAR MODULAR EQUATIONS}
A x=
\begin{pmatrix}
a_1(1) & a_2(1) & \cdots & a_n(1)\\
a_1(2) & a_2(2) & \cdots & a_n(2)\\
\vdots & \vdots & \cdots & \vdots\\
a_1(m) & a_2(m) & \cdots & a_n(m)
\end{pmatrix}
\begin{pmatrix}
x(1)\\
x(2)\\
\vdots\\
x(n)
\end{pmatrix}
=
\begin{pmatrix}
b(1) \\
b(2) \\
\vdots\\
b(m)
\end{pmatrix}
\begin{matrix}
\mod d_1\\
\mod d_2\\
\vdots\\
\mod d_m
\end{matrix}
= b \pmod G
\end{equation}
We will refer to $A$ and to (\ref{INITIAL SYSTEM OF LINEAR MODULAR EQUATIONS}) as our \textit{original matrix} and our \textit{original system} of equations, the latter we aim to solve. Now, let us denote by $d$ the least common multiplier of $c_1,\ldots,c_n, d_1, \ldots, d_m$; note that $d$ is upper bounded by $|H||G|$. To solve (\ref{INITIAL SYSTEM OF LINEAR MODULAR EQUATIONS}), our approach will be, first, to transform (\ref{INITIAL SYSTEM OF LINEAR MODULAR EQUATIONS})  into a slightly-larger system of congruences modulo $d$; second, to apply methods from computational number theory  to deal with the latter.

\subsection{Structure of the solutions and initial simplifications}

 Let $\left(x_0, x_1,\ldots,x_r\right)$  be an  $r$-tuple of elements of $H$ such that $x_0$ is a particular solution of (\ref{INITIAL SYSTEM OF LINEAR MODULAR EQUATIONS}) and $ x_1,\ldots,x_r$ generate $\ker{A}$; any tuple like $\left(x_0, x_1,\ldots,x_r\right)$ will be called a \textit{general solution} since, due to (\ref{Structure of the solutions of A(x)=b EQUATION}), the set $X_{sol}$ of solutions  of (\ref{INITIAL SYSTEM OF LINEAR MODULAR EQUATIONS}) is spanned by linear combinations of the form $x_0 + \sum k_i x_i$. We will not impose restrictions on the integer $r$ apart that it must be $O(\polylog{|H|})$. Using this notation, the system (\ref{INITIAL SYSTEM OF LINEAR MODULAR EQUATIONS}) is unsolvable if and only if its (unique) general solution is the empty tuple $()$.

We will now argue (in two steps), that the group where solutions must live, $H$, can be chosen w.l.o.g to be $\Z_d^{n}$,  a simplification that will be exploited in subsequent sections. This is good enough for our purposes, since the order of this group is $O(\polylog{|G|,|H|})$:\\

First, using (\ref{Homomorphism condition}) and the fact that $dg=0$ for all elements $g\in G$, one readily sees that $A$ defines a homomorphism $\mathcal{A}$ from $\Z_d^{n}$ to $G$ \footnote{The new symbol is used to distinguish $\mathcal{A}$ from the original homomorphism $A:H\rightarrow G$.}; therefore, the system $Ax=b\pmod{G}$ with $x\in \Z_d^{n}$ is linear and its solutions (if there are any) form a coset $\mathcal{X}_{sol}={x}_0+\ker\mathcal{A}$.\\

Second, remark that $H$ is a subset of $\Z_d^{n}$ and that every $c_i$ divides $d$; as a result, the projection $\pi({x})={x}\pmod H$ is seen to be a (surjective) group homomorphism $\Z_d^{n}\rightarrow H$ using (\ref{Homomorphism condition}). Now we show that to solve (\ref{INITIAL SYSTEM OF LINEAR MODULAR EQUATIONS}) we can, first, look for solutions  $x$ inside the bigger space $\Z_d^{n}$ and, second, project them onto $H$ using $\pi({x})$. On the one hand, it is easy to see that for every  solution $x\in \Z_d^{n}$ the projected solution $\pi(x)$ is also a solution: using $x(i)=\pi({x})(i)+kc_i$ we derive $\pi(x)=x-\sum k_ic_ie_i$ and
\begin{equation}
\label{projected solutions EQUATION} A\pi({x})=A{x}-\sum k_i c_i a_i=A{x}=b\pmod{G}.
\end{equation}
In (\ref{projected solutions EQUATION}) we used (\ref{Homomorphism condition}) in the second equality. On the other hand, since we search for solutions inside a set larger than $H$ and $\pi(x)=x$ for every solution $x\in H$, it follows
\begin{equation}\label{projecting solution sets EQUATION}
X_{sol}=\pi(\mathcal{X}_{sol})=\{ \pi(x):\: Ax=b\pmod{G}\,\,, x\in \Z_d^{n}\}
\end{equation}
Now, imagine we are given a general solution of the system $(x_0, x_1,\ldots, x_r)$ where all $x_i\in \Z_d^{n}$. In view of above properties, it follows that the projected tuple $(\pi(x_0), \pi(x_1),\ldots, \pi(x_r))$ is a general solution of (\ref{INITIAL SYSTEM OF LINEAR MODULAR EQUATIONS}); were the former general solution empty, we would conclude that (\ref{INITIAL SYSTEM OF LINEAR MODULAR EQUATIONS}) admits no solution; and, moreover, projecting with $\pi$ reduces the number of distinct solutions by a multiplicative factor of $|\ker{\pi}|=|\langle c_1 \rangle\times \cdots\times \langle c_n \rangle|=d^{n}/(c_1 c_2\cdots c_n)$---to see this, we choose $b=0$ in (\ref{projecting solution sets EQUATION}), which leads to $\ker A= \text{im} \pi_{|\mc{X}_{sol}} \cong \ker \mathcal{A}/\ker{\pi_{|\mc{X}_{sol}}}$.

\subsection{Enlarging (\ref{INITIAL SYSTEM OF LINEAR MODULAR EQUATIONS}) to a system of linear congruences}

 In this section we show how to reduce our original system (\ref{INITIAL SYSTEM OF LINEAR MODULAR EQUATIONS}) to a system of linear congruences, assuming  that $H=\Z_d^{n}$. The first steps are rather standard: first, we `undo' the modular equations of the initial system (\ref{INITIAL SYSTEM OF LINEAR MODULAR EQUATIONS})  introducing $m$ new integer variables and take remainders modulo $d$; the final system of equations will be denoted the \textit{enlarged system}  associated to (\ref{INITIAL SYSTEM OF LINEAR MODULAR EQUATIONS}).
\begin{equation}\label{Table: initial and enlarged system of equations}
\begin{matrix}
\textnormal{\underline{Original system}}& \quad & \textnormal{\underline{Enlarged system}}\\\noalign{\medskip}
Ax = b \pmod G   & \longrightarrow & \boldsymbol{A}\boldsymbol{x} =
\begin{pmatrix}
A&D
\end{pmatrix}
\begin{pmatrix}
\boldsymbol{{x}}_1\\
\boldsymbol{x}_2
\end{pmatrix}
=b
\mod d \\\noalign{\smallskip}
x\in H=\Z_d^{n} & & \boldsymbol{x}\in\Z_d^{n}\times  \Z_d^{m}
\end{matrix}
\end{equation}
Above, $D=\textnormal{diag}(d_1,\ldots, d_m)$ is a diagonal $m\times m$ integer matrix and, with little abuse of notation, we embedded the columns $a_j$ of $A$ and $b$ (originally elements of $G$) inside the larger group $\Z_d^{m}$. Since the former group is a subset of the latter, this operation can be implemented via a rudimentary inclusion map from $G$ to $\Z_d^{m}$.

It is easy to check that the enlarged matrix $\boldsymbol{A}$ defines a group homomorphism from $\Z_d^{n}\times  \Z_d^{m}$ to $\Z_d^{m}$, using (\ref{Homomorphism condition}). Therefore,  the enlarged system is a system of linear congruences modulo $d$. It follows that the sets of solutions of the initial and the enlarged system  (\ref{Table: initial and enlarged system of equations}), respectively denoted by  $X_{sol}$ and $\boldsymbol{X}_{sol}$, have a coset structure
\begin{equation}\label{Coset structure of the Set of Paritcular Solutions EQUATION}
X_{sol}=x_0+\ker{A},  \qquad\qquad\qquad \quad \boldsymbol{X}_{sol}=\boldsymbol{x}_0+\ker{\boldsymbol{A}}.
\end{equation}
The following lemma shows that the solutions of the original and the enlarged system are intrinsically related, and gives a $\polylog(|G|,|H|)$ reduction of the former into the latter.
\begin{lemma}\label{REDUCTION FROM ORIGINAL TO ENLARGED SYSTEM}
Let $X_{sol}$ and $\boldsymbol{X}_{sol}$ be, respectively, the sets of solutions of the original and enlarged systems of linear congruences shown in (\ref{Table: initial and enlarged system of equations}). Then the following propositions hold:
\begin{enumerate}
\item[(a)] The original system admits  solutions iff the enlarged system admits  solutions.
\item[(b)] The solutions of the original system can be obtained from those of the enlarged system via
\begin{equation}\label{enlarged System can be projected to Initial System EQUATION}
X_{sol}=\pi(\boldsymbol{X}_{sol}),
\end{equation}
where $\pi(\boldsymbol{x}):=(\boldsymbol{x}(1),\ldots,\boldsymbol{x}(n))$  is a surjective group homomorphism  from $\Z_d^{n+m}$ to $\Z_d^{n}$.
\item[(c)] The cardinality of both sets are related through $|\boldsymbol{X}_{sol}|=|G||X_{sol}|$.
\end{enumerate}
\end{lemma}
Notice that a direct consequence of lemma \ref{REDUCTION FROM ORIGINAL TO ENLARGED SYSTEM} is that we can efficiently (i) \textbf{decide} whether the \textit{original} system admits solutions, (ii) \textbf{find} a general-solution for it and (iii) \textbf{count} its number of solutions \textit{if} an efficient subroutine to solve these problems for the \textit{enlarged} system is provided: in particular, given a general-solution $\left( \boldsymbol{x}_0, \ldots, \boldsymbol{x}_r \right)$ of (\ref{Table: initial and enlarged system of equations}), it follows that $\left(\pi(\boldsymbol{x}_0),\ldots,\pi(\boldsymbol{x}_r)\right)$ is a general-solution of (\ref{INITIAL SYSTEM OF LINEAR MODULAR EQUATIONS}).

The next theorem states, furthermore, that there are efficient classical algorithms to solve systems of linear equations modulo $d$.
\begin{theorem}\label{THM SOLVING SYSTEMS OF LINEAR CONGRUENCES}
Given a system of linear congruences $\boldsymbol{A}\boldsymbol{x}=b\bmod{d}$ where $\boldsymbol{A}\in \Z_d^{m\times n}$, $b\in\Z_d^{m}$ and $\boldsymbol{x}\in \Z_d^{n}$, there exist deterministic $\ppoly{m,n, \log{d}}$ classical algorithms to solve the following tasks: (a) deciding whether the system admits solutions; (b) computing a general solution; (c) counting the number of different solutions.
\end{theorem}
Given lemma \ref{REDUCTION FROM ORIGINAL TO ENLARGED SYSTEM} and theorem \ref{THM SOLVING SYSTEMS OF LINEAR CONGRUENCES} (both to be proven below), the proof of theorem \ref{THM SYSTEM OF LINEAR CONGRUENCES MODULO AN ABELIAN GROUP} is completed; the remaining two sections are devoted to prove these last two results.

\subsection{Proof of lemma \ref{REDUCTION FROM ORIGINAL TO ENLARGED SYSTEM}}

We will first prove a smaller result.
\begin{lemma}[\textbf{Kernel of a diagonal matrix}]\label{Kernel of a diagonal matrix}
Given $S\in\Z_d^{m\times n}$, a diagonal matrix whose $r$-first diagonal coefficients $s_1,\ldots, s_r$ are positive integers and the rest are equal to zero; let $S:\Z_d^{n}\rightarrow\Z_d^{m}$ be the group-homomorphism defined by $S$ via matrix multiplication; then, the kernel of $S$ fulfills
\begin{equation}
\ker{S}=\langle \widetilde{s_1} \rangle\times\cdots\times\langle \widetilde{s_r} \rangle\times \Z_{d}^{n-r} \quad \text{and} \quad |\ker S| = q_1 q_2 \cdots q_r \, d^{n-r},
\end{equation}
where $q_i:=gcd(s_i,d)$ and $\widetilde{s_i}= d/q_i$.
\end{lemma}
\begin{proof}
Any $x$ fulfilling $Sx=0\mod d$ must satisfy $r$ constrains $s_i x(i)=k_i d$, for $i\leq r$,  where  $k_i$ are arbitrary integers. If we divide both sides by the $q_i$, we can derive a new set of equivalent modular constrains
\begin{equation*}
\frac{s_i}{q_i}x(i)=0\mod{\frac{d}{q_i}}, \quad \textnormal{for all}\;   i\leq r
\end{equation*}
Now, each number $(s_i/q_i)$ is coprime to $d_i/q_i$; hence, each $(s_i/q_i)$ has an inverse element in $\Z_{d_i/q_i}$ and can be removed from the constrain where it appears multiplying it by the latter. It follows that the possible values for $x(i)$ are the multiples of $\widetilde{s_i}=d/q_i$ inside $\Z_d$, proving the first equation; the second equation, follows as a consequence, for $|d/q_i|= q_i$.\end{proof}

\noindent \textbf{We prove now lemma \ref{REDUCTION FROM ORIGINAL TO ENLARGED SYSTEM}(b), which also implies (a):} For every $x\in X_{sol}$ we can define a tuple $y_x\in\Z_d^{m}$ coefficient-wise as
\begin{equation}
y_x(i):= \left[b-Ax\right](i)/d_i \mod d.
\end{equation}
It follows easily that all elements with the form $\boldsymbol{{x}}=\binom{x}{y_x}$ where $x\in X_{sol}$ belong to $\boldsymbol{X}_{sol}$, by checking $[A|D]\boldsymbol{x}=b\bmod d$. Moreover each $\boldsymbol{x}$  satisfies $x=\pi(\boldsymbol{x})$ and, thence, it follows that $X_{sol}\subseteq \pi(\boldsymbol{X}_{sol})$. We finish the proof of (b) showing that the inclusion $X_{sol}\supseteq \pi(\boldsymbol{X}_{sol})$ also holds. The strategy is now to prove that for every $\boldsymbol{x}=\binom{x}{y}\in \boldsymbol{X}_{sol}$ the element $x=\pi(\boldsymbol{x})$ belongs to $X_{sol}$, which can be shown as follows:
\begin{equation}
\boldsymbol{A}\boldsymbol{x}=Ax+Dy = b \mod d \Rightarrow Ax(i)=b(i)-y(i)d_i + k_id \quad  \textnormal{for all $i=1,\ldots,m$}.
\end{equation}
Since $d_i$ divides $d$, the last two terms are 0 modulo $d_i$. Thus, $Ax(i)=b(i)\mod d_i$ for every $i$, and we are done. \qed\\

\noindent \textbf{Finally, we prove lemma \ref{REDUCTION FROM ORIGINAL TO ENLARGED SYSTEM}(c):} Every pair elements of $\boldsymbol{X}_{sol}$ that are mapped to $x$ via $\pi$ must have the form $\boldsymbol{x}_1=\binom{x}{y_1}$, $\boldsymbol{x}_2=\binom{x}{y_2}$ where $D(y_1-y_2)=0\mod d$. Therefore, exactly $|\ker D|$ distinct elements of $\boldsymbol{X}_{sol}$ are mapped to $x$, for any  value of $x$. Applying lemma \ref{Kernel of a diagonal matrix} to $D$, it follows $|\boldsymbol{X}_\text{sol}|=|\ker D||{X}_\text{sol}|=d_1 d_2 \cdots d_m |{X}_\text{sol}|$, as desired. \qed

\subsection{Proof of theorem \ref{THM SOLVING SYSTEMS OF LINEAR CONGRUENCES}}

 We will use existing algorithms to compute the Smith normal form of $A$ over the integers modulo $d$ \cite{storjohann10_phd_thesis}: these return an $m\times m$ matrix $U$ and an $n\times n$ matrix $V$, both \textit{invertible}, such that $S:=U\boldsymbol{A}V$ is diagonal and its first $r$ diagonal coefficients are non-zero (with $r$ depending of the particular problem). Instead of $\boldsymbol{A}\boldsymbol{x}=b\mod d$ we  will solve the system $S\boldsymbol{y}=c\mod d$ with  $c=Ub\mod d$: since $V$ is an invertible homomorphism, it follows that $\boldsymbol{y}_0$ is a solution of this system iff $V\boldsymbol{y}_0$ is a solution of the initial system; owing to, if we solve (a-b-c) for the new ``diagonal system" we are done.

Finally, recall that we can use the extended Euclidean algorithm to \textit{decide} the solvability of any linear congruence $s_i\boldsymbol{y}(i)=c(i)\bmod d$, and find (if it exists) a particular solution $\boldsymbol{y}_0$. These facts, together with lemma \ref{Kernel of a diagonal matrix}, (which shows how to obtain a generating-set of $\ker{S}$), let us compute a \textit{general-solution} of $S\boldsymbol{y}=c\mod d$.  Moreover, the \textit{number} of solutions of the system---either $0$ or $|\ker{S}|$---can also be computed due to lemma \ref{Kernel of a diagonal matrix}.

\end{document}